\documentclass[aps, prd, showpacs, preprintnumbers, floatfix,nofootinbib,longbibliography,superscriptaddress,reprint]{revtex4-2}
\usepackage{graphicx}
\usepackage{dcolumn}
\usepackage{bm}
\usepackage{hyperref}
\usepackage[mathlines]{lineno}
\usepackage[dvipsnames]{xcolor}
\usepackage{acronym}
\usepackage{physics}
\usepackage{mathtools}
\usepackage{longtable}
\usepackage{booktabs}
\usepackage{multirow}
\usepackage{placeins}
\usepackage{float}
\usepackage{orcidlink}
\usepackage{ifthen}
\usepackage{etoolbox}
\usepackage{siunitx}
\usepackage[normalem]{ulem}

\newcommand{\tabref}[1]{Tab.~\ref{#1}}
\renewcommand{\eqref}[1]{Eq.~(\ref{#1})}
\newcommand{\eqsref}[2]{Eqs.~(\ref{#1}) and (\ref{#2})}
\newcommand{\figref}[1]{Fig.~\ref{#1}}

\newcommand{\secref}[1]{Sec.~\ref{#1}}
\newcommand{\appref}[1]{Appendix~\ref{#1}}
\newcommand{\appsref}[2]{Appendices~\ref{#1} and \ref{#2}}
\newcommand{\Ogw}{\Omega_\text{GW}}
\newcommand{\calP}{\mathcal{P}}

\newcommand{\phat}{\hat{\calP}}

\newcommand{\lmax}{\ell_\text{max}}

\begin{document}
\include{latex_macros}

\preprint{LIGO-P250038}

\title{Directional Search for Persistent Gravitational Waves: Results from the First Part of LIGO-Virgo-KAGRA's Fourth Observing Run}

\author{A.~G.~Abac\,\orcidlink{0000-0003-4786-2698}}
\affiliation{Max Planck Institute for Gravitational Physics (Albert Einstein Institute), D-14476 Potsdam, Germany}
\author{I.~Abouelfettouh}
\affiliation{LIGO Hanford Observatory, Richland, WA 99352, USA}
\author{F.~Acernese}
\affiliation{Dipartimento di Farmacia, Universit\`a di Salerno, I-84084 Fisciano, Salerno, Italy}
\affiliation{INFN, Sezione di Napoli, I-80126 Napoli, Italy}
\author{K.~Ackley\,\orcidlink{0000-0002-8648-0767}}
\affiliation{University of Warwick, Coventry CV4 7AL, United Kingdom}
\author{C.~Adamcewicz\,\orcidlink{0000-0001-5525-6255}}
\affiliation{OzGrav, School of Physics \& Astronomy, Monash University, Clayton 3800, Victoria, Australia}
\author{S.~Adhicary\,\orcidlink{0009-0004-2101-5428}}
\affiliation{The Pennsylvania State University, University Park, PA 16802, USA}
\author{D.~Adhikari}
\affiliation{Max Planck Institute for Gravitational Physics (Albert Einstein Institute), D-30167 Hannover, Germany}
\affiliation{Leibniz Universit\"{a}t Hannover, D-30167 Hannover, Germany}
\author{N.~Adhikari\,\orcidlink{0000-0002-4559-8427}}
\affiliation{University of Wisconsin-Milwaukee, Milwaukee, WI 53201, USA}
\author{R.~X.~Adhikari\,\orcidlink{0000-0002-5731-5076}}
\affiliation{LIGO Laboratory, California Institute of Technology, Pasadena, CA 91125, USA}
\author{V.~K.~Adkins}
\affiliation{Louisiana State University, Baton Rouge, LA 70803, USA}
% ...


\begin{abstract}
The angular distribution of gravitational-wave power from persistent sources may exhibit anisotropies arising from the large-scale structure of the Universe. This motivates directional searches for astrophysical and cosmological gravitational-wave backgrounds, as well as continuous-wave emitters.
We present results of such a search using data from the first observing run through the first portion of the fourth observing run of the LIGO-Virgo-KAGRA Collaborations. We apply gravitational-wave radiometer techniques to generate skymaps and search for both narrowband and broadband persistent gravitational-wave sources. Additionally, we use spherical harmonic decomposition to probe spatially extended sources. 
No evidence of persistent gravitational-wave signals is found, and we set the most stringent constraints to date on such emissions. For narrowband point sources, our sensitivity estimate to effective strain amplitude lies in the range $(0.03 - 8.4) \times 10^{-24}$ across all sky and frequency range $(20 - 160)$ Hz. For targeted sources---Scorpius X-1, SN 1987A, the Galactic Center, Terzan 5, and NGC 6397---we constrain the strain amplitude with best limits ranging from $\sim 1.1 \times 10^{-25}$ to $6.5 \times 10^{-24}$. For persistent broadband sources, we constrain the gravitational-wave flux $F_{\alpha, \hat{n}}^{95\%, \mathrm{UL}}(25\, \mathrm{Hz}) < (0.008 - 5.5) \times 10^{-8}\, \mathrm{erg\, cm^{-2}\, s^{-1}\, Hz^{-1}}$, depending on the sky direction $\hat{n}$ and spectral index $\alpha=0,\,2/3,\,3$. Finally, for extended sources, we place upper limits on the strain angular power spectrum $C_\ell^{1/2} < (0.63 - 17) \times 10^{-10} \,\mathrm{sr}^{-1}$.
\end{abstract}

\maketitle

\acrodef{LSC}[LSC]{LIGO Scientific Collaboration}
\acrodef{LVC}[LVC]{LIGO Scientific and Virgo Collaborations}
\acrodef{LVK}[LVK]{LIGO-Virgo-KAGRA Collaboration}
\acrodef{aLIGO}{Advanced Laser Interferometer Gravitational-Wave Observatory}
\acrodef{aVirgo}{Advanced Virgo}
\acrodef{LIGO}[LIGO]{Laser Interferometer Gravitational-Wave Observatory}
\acrodef{IFO}[IFO]{interferometer}
\acrodef{LHO}[LHO]{LIGO-Hanford}
\acrodef{LLO}[LLO]{LIGO-Livingston}
\acrodef{O2}[O2]{second observing run}
\acrodef{O1}[O1]{first observing run}
\acrodef{O3}[O3]{third observing run}
\acrodef{O3a}[O3a]{first half of the third observing run}
\acrodef{O3b}[O3b]{second half of the third observing run}
\acrodef{O4}[O4]{fourth observing run}
\acrodef{O4a}[O4a]{the first portion of the fourth observing run}

\acrodef{BH}[BH]{black hole}
\acrodef{BBH}[BBH]{binary black hole}
\acrodef{BNS}[BNS]{binary neutron star}
\acrodef{IMBH}[IMBH]{intermediate-mass black hole}
\acrodef{NS}[NS]{neutron star}
\acrodef{BHNS}[BHNS]{black hole--neutron star binaries}
\acrodef{NSBH}[NSBH]{neutron star--black hole binary}
\acrodef{PBH}[PBH]{primordial black hole binaries}
\acrodef{CBC}[CBC]{compact binary coalescence}
\acrodef{GW}[GW]{gravitational wave}
\acrodef{CGW}[CGW]{continuous gravitational wave}
\acrodef{CW}[CW]{continuous waves}
\acrodef{GWH}[GW]{gravitational-wave}
\acrodef{GWB}[GWB]{gravitational-wave background}
\acrodef{UL}[UL]{upper limit}

\acrodef{CWB}[cWB]{coherent WaveBurst}

\acrodef{SNR}[SNR]{signal-to-noise ratio}
\acrodef{FAR}[FAR]{false alarm rate}
\acrodef{IFAR}[IFAR]{inverse false alarm rate}
\acrodef{FAP}[FAP]{false alarm probability}
\acrodef{PSD}[PSD]{power spectral density}
\acrodef{CSD}[CSD]{cross spectral density}

\acrodef{GR}[GR]{general relativity}
\acrodef{NR}[NR]{numerical relativity}
\acrodef{PN}[PN]{post-Newtonian}
\acrodef{EOB}[EOB]{effective-one-body}
\acrodef{ROM}[ROM]{reduced-order model}
\acrodef{IMR}[IMR]{inspiral--merger--ringdown}

\acrodef{PDF}[PDF]{probability density function}
\acrodef{PE}[PE]{parameter estimation}
\acrodef{CL}[CL]{credible level}

\acrodef{KLD}[KLD]{Kullback--Leibler divergence}
\acrodef{JSD}[JSD]{Jensen--Shannon divergence}

\acrodef{BPL}[BPL]{broken power law}
\acrodef{PL}[PL]{power law}
\acrodef{RSS}[RSS]{residual sum of squares}
\acrodef{ORF}[ORF]{overlap reduction function}
\acrodef{PTA}[PTA]{pulsar timing array}

\acrodef{ASAF}[ASAF]{all-sky-all-frequency radiometer}
\acrodef{BBR}[BBR]{broadband radiometer}
\acrodef{NBR}[NBR]{narrowband radiometer}
\acrodef{SPH}[SPH]{spherical harmonics}
\acrodef{SHD}[SHD]{spherical harmonic decomposition}

\acrodef{SFT}[SFT]{short Fourier transform}

\acrodef{LMXB}[LMXB]{low mass X-ray binary}

\newcommand{\PN}[0]{\ac{PN}\xspace}
\newcommand{\BBH}[0]{\ac{BBH}\xspace}
\newcommand{\BNS}[0]{\ac{BNS}\xspace}
\newcommand{\BH}[0]{\ac{BH}\xspace}
\newcommand{\NR}[0]{\ac{NR}\xspace}
\newcommand{\SNR}[0]{\ac{SNR}\xspace}
\newcommand{\aLIGO}[0]{\ac{aLIGO}\xspace}
\newcommand{\PE}[0]{\ac{PE}\xspace}
\newcommand{\IMR}[0]{\ac{IMR}\xspace}
\newcommand{\PDF}[0]{\ac{PDF}\xspace}
\newcommand{\GR}[0]{\ac{GR}\xspace}
\newcommand{\PSD}[0]{\ac{PSD}\xspace}
\newcommand{\EOS}[0]{\ac{EOS}\xspace}
\newcommand{\LVC}[0]{\ac{LVC}\xspace}

\newcommand{\pystoch}{\texttt{PyStoch}}
\newcommand{\pygwb}{\tt{pyGWB}\xspace}

\newcommand{\PYTHON}{\texttt{Python}\xspace}
\newcommand{\NUMPY}{\texttt{NumPy}\xspace}
\newcommand{\SCIPY}{\texttt{SciPy}\xspace}
\newcommand{\PLT}{\texttt{Matplotlib}\xspace}
\newcommand{\SEABORN}{\texttt{seaborn}\xspace}
\newcommand{\GWPY}{\texttt{GWpy}\xspace}
\newcommand{\DYNESTY}{\texttt{Dynesty}\xspace}

\section{Introduction}
\label{sec:introduction}
A \ac{GWB} is a diffuse signal resulting from the incoherent superposition of numerous unresolved \ac{GW} sources. 
Its stochastic nature may stem from either the formation mechanisms of the sources or the limited sensitivity of current detectors. 
A wide range of sources is expected to contribute to the \ac{GWB}, each with distinct characteristics and frequency signatures. The \ac{GWB} is typically categorized into two main types: the astrophysical \ac{GWB} \cite{Regimbau:2011rp}, arising from unresolved sources such as compact binary coalescences \cite{CBC_Wu:2011ac, CBC_Zhu:2011bd, CBC_Zhu:2012xw, CBC_Rosado:2011kv, CBC_Marassi:2011si} and rotating neutron stars \cite{core_collapse_Crocker:2015taa, core_collapse_Crocker:2017agi, core_collapse_Ferrari:1998ut, magnetars_2013PhRvD..87d2002W, magnetars_Marassi:2010wj, f-modes_Lasky:2013jfa, r-modes_Ferrari:1998jf, r-modes_Zhu:2011pt, isolated_NSs_magnetars_Regimbau:2001kx}, and the cosmological \ac{GWB} \cite{Caprini-Figueroa:2018mtu}, 
potentially generated by early-universe phenomena
such as inflation \cite{inflation_1979JETPL, inflation_Bar-Kana:1994nri, inflation_Turner:1996ck}, cosmic strings \cite{cosmic_strings_Damour:2004kw, cosmic_strings_Kibble:1976sj, cosmic_strings_Sarangi:2002yt, cosmic_strings_Siemens:2006yp}, or phase transitions \cite{FOPT_Marzola:2017jzl, FOPT_VonHarling:2019rgb}.

While to first order the \ac{GWB} is expected to be homogeneous and isotropic across the sky, it may exhibit two kinds of anisotropy: large-scale and local. 
Large-scale anisotropies may arise from the uneven distribution of sources on cosmological scales~\cite{Jenkins:2019nks, Jenkins:2019uzp, Alonso:2020mva, Cusin:2019jpv, Yang:2020usq, Yang:2023eqi, Cusin:2018rsq, Capurri:2021zli, Alonso:2024knf}, propagation effects due to large-scale structures along the line of sight~\cite{Bertacca:2019fnt}, and kinematic anisotropies caused by the motion of the observer relative to the \ac{GWB} rest frame~\cite{Allen:1996gp,Cusin:2022cbb,ValbusaDallArmi:2022htu,Chung:2022xhv,kinematic_boosting_correction}. In contrast, local anisotropies can be caused by nearby, spatially clustered sources, for example a concentration of millisecond pulsars in regions such as the Galactic plane and Virgo cluster, leading to prominent hotspots in the sky~\cite{Allen:1996gp,Dhurandhar:2011qe,Mazumder:2014fja,Agarwal:2022lvk,DeLillo:2022blw, PhysRevD.89.123008}.

The \ac{LVK} has set progressively improved \acp{UL} on the \ac{GWB} energy density of both isotropic and anisotropic components \cite{LIGOScientific:2007gwp,LIGOScientific:2011kvu, LIGOScientific:2016jlg, LIGOScientific:2019vic, KAGRA:2021kbb, LIGOScientific:2016nwa,LIGOScientific:2019gaw,KAGRA:2021mth,KAGRA:2021rmt}. 
To look for \ac{GWB} anisotropies, the \ac{LVK} performs directional searches using a \ac{GW}-radiometer algorithm to generate skymaps and a \ac{SPH} decomposition to compute angular power spectra \cite{Ballmer:2005uw, Mitra:2007mc, PhysRevD.80.122002}.
These tools can be used to identify cosmological, astrophysical, or local anisotropies. 
In addition, the \ac{GW} radiometer algorithm designed to search for \ac{GWB} is also well suited to look for other types of persistent GW sources, including \ac{CGW} sources.

Until the \ac{O3}, the pixel-based anisotropic searches \cite{LIGOScientific:2016nwa,LIGOScientific:2019gaw,KAGRA:2021mth} performed by \ac{LVK}, targeting point-like persistent sources, typically followed two approaches: 
\ac{NBR} searches, which focused on specific directions (Scorpius X-1,
the Galactic Center, SN 1987A, etc.) across multiple narrow frequency bins, and \ac{BBR} searches, which scanned the entire sky but averaged over a wide frequency range. As a result, the prospects of detecting unknown anisotropies were limited, since neither approach could simultaneously explore the full angular and spectral properties of the signal. In addition, matched-filtering-based searches for persistent signals from galactic or extragalactic sources, such as neutron stars \cite{Wette:2023dom,Piccinni:2022vsd,Riles:2022wwz} or boson clouds around black holes \cite{LIGOScientific:2021rnv, Jones:2023fzz, LIGOScientific:2025csr}, are computationally expensive and inherently limited by signal modeling assumptions. To address this limitation, we introduced a new strategy in \ac{O3}: performing directional searches in narrow frequency bins across the whole sky. This approach, formalized as the \ac{ASAF} search \cite{KAGRA:2021rmt}, enables a more comprehensive exploration of potential anisotropic signals without prior assumptions on their location or frequency content.

All of the above searches are performed in pixel basis, which is well-suited for probing localized sources. However, as mentioned above, \ac{GWB} anisotropies can also be expanded in the \ac{SPH} basis, which is more appropriate for studying extended or diffuse sources. The \ac{LVK} has also conducted searches using the SPH basis and reported \acp{UL} on the angular power spectra using data up to \ac{O3} \cite{LIGOScientific:2016nwa,LIGOScientific:2019gaw,KAGRA:2021mth}. 

In this paper, we present results from all four analyses: \ac{ASAF}, targeted \ac{NBR}, \ac{BBR}, and \ac{SPH}, performed on the \ac{LVK} observational data from the \ac{O1}, \ac{O2}, \ac{O3}, and \ac{O4a}. These analyses benefit from data collected by an increasingly sensitive global network of \ac{GW} detectors, whose improved sensitivity over the years has significantly enhanced our ability to probe anisotropic sources of persistent \acp{GW}. Despite this progress, we do not find evidence for such
sources in any of the four analyses and therefore set \acp{UL} on the \ac{GW} emission depending on specific sky directions or angular scales.

We note that this paper has two companions, one focusing on new results of the isotropic \ac{GWB} search \cite{O4a_ISO} and the other discussing cosmological implications of the new isotropic search results \cite{O4a_CosmoImpli}.

The paper is organized as follows. In \secref{sec:methods}, we introduce the \ac{GW} radiometer algorithm and search methodologies. In \secref{sec:Results}, we present results from \ac{ASAF} radiometer search, targeted \ac{NBR} search, \ac{BBR} search, and \ac{SPH} search for extended sources. Finally, in \secref{sec:conclusion}, we summarize our findings and outline prospects for future searches. Comprehensive descriptions of the individual analyses can be found in the Appendices. 

\section{Motivation and Methods}
\label{sec:methods}
\subsection{Gravitational Wave Radiometer}
\label{sec:general_methods}
The dimensionless GW energy density parameter, $\Omega_{\text{GW}}(f,\,\hat{n})$, characterizes the energy content of the \ac{GWB} per unit logarithmic frequency $f$ and unit solid angle in the direction $\hat{n}$. It is given by
\begin{align}
     \Omega_{\text{GW}}(f,\,\hat{n}) = \frac{2 \pi^2}{3 H_{0}^2}\, f^3 \,\mathcal{P}(f,\,\hat{n})\,,
\end{align}
where $\mathcal{P}(f,\hat{n})$ denotes the one-sided \ac{PSD} of the GW strain field. This quantity corresponds to the second moment of a stationary, Gaussian, and unpolarized stochastic GW signal~\cite{sgwb_unified_treatment}. $H_{0} = 67.9\,\mathrm{km\, s^{-1} Mpc^{-1}}$ in the above equation denotes the Hubble constant~\cite{Planck:2015fie}. 

We aim to measure the angular distribution of the \ac{GWB}, i.e., $\mathcal{P}(f,\hat{n})$, by performing a cross-correlation\footnote{We note that the estimator constructed via cross-correlation is nearly optimal, given that the auto-correlation is not utilized to detect the GWB signal~\cite{KAGRA:2021kbb}.} between data from a pair of detectors (usually referred to as baseline, denoted by $I$, with the subscripts $i = 1,2$ labeling the two detectors) that are geographically separated.
We construct the \ac{CSD} estimator:
\begin{equation}
    C^{I}(t;f)\equiv\frac{2}{T\, \overline{w_1 w_2}} \tilde{s}^*_{1}(t;f)\,\tilde{s}_{2}(t;f)\,,
\end{equation}
where $\tilde{s}_{i}(t;f)$ represents the \ac{SFT} computed from a time segment centered around time $t$, using a $50\%$ overlapping Hann window with a segment duration of $T=192$ s. The factor $\overline{w_1 w_2}$ accounts for the effect of windowing on the estimator~\cite{albert_joe,Ain:2015lea}.

In the presence of Gaussian, additive, and stationary noise---assumed to be uncorrelated between detectors at different sites---the noise- and source-averaged correlation, denoted by $\langle\cdot \rangle_{\tilde{h},N}$, is given by 
\begin{equation}
   \langle  C^{I}(t;f) \rangle_{\tilde{h},N}=\int d^2\hat{n}\,\gamma^I(t;f,\hat{n})\, \mathcal{P}(f,\hat{n})\,,
\end{equation}
which encodes information about the source anisotropy. The \ac{ORF} \cite{10.1093/mnras/227.4.933,PhysRevD.46.5250,Flanagan:1993ix}, $\gamma^I(t;f,\hat{n})$, acts as a transfer function by mapping the spatial distribution of \ac{GW} power onto the measured cross-correlation. It depends on the signal frequency, the observation time, the source sky location, the individual detector response functions, and the distance between the detectors. 
\begin{widetext}
To probe specific types of angular distributions, we discretize the sky $\mathcal{P}(f,\hat{n})$ using an appropriate basis $e_\mu(\hat{n})$, i.e.,
\begin{equation}
    \mathcal{P}(f,\hat{n}) \equiv \sum_\mu \mathcal{P}_\mu(f)\,e_\mu(\hat{n})\, .
\end{equation}
The explicit expression for the \ac{ORF} in an appropriate basis is given by
\begin{equation}\label{eq:ORF}
    \gamma^I_\mu(t;f) \equiv \sum_{A=+,\times}\int d^2\hat{n}\, F^A_{1}(t,\hat{n}) F^A_{2}(t,\hat{n}) \,e_\mu(\hat{n})\,{\rm e}^{-2\pi i f  \frac{\Delta\vec{x}^I(t)\cdot \hat{n}}{c}}\, ,
\end{equation}
where $F^A_{i}$ denotes the individual detector response functions, and $\Delta\vec{x}^I$ is the separation vector between the detectors.
\end{widetext}
In this article, we utilize two types of bases: (i) the pixel basis $e_\mu(\hat{n})=\delta^2(\hat{n}-\hat{n}_\mu)$ under the assumption the source is localized in the direction $\hat{n}_\mu$; and (ii) the \ac{SPH} basis $e_\mu(\hat{n})=Y_{\ell m}(\hat{n})$, with $\mu=(\ell,m)$, under the assumption the source is spatially extended. We note that the estimators can be transformed between pixel and \ac{SPH} basis via forward and inverse \ac{SPH} transforms~\cite{Suresh:2020khz}. Additionally, it is important to note that
$e_\mu(\hat{n})=Y_{00}(\hat{n})$ corresponds to the isotropic component of the GWB angular distribution.

\begin{widetext}
Assuming that the \ac{CSD} for each time segment is a Gaussian random variable, the maximum likelihood estimator for \ac{GWB} anisotropy, $\mathcal{P}_{\mu}(f)$, (for positive frequency) is obtained by maximizing the joint likelihood across different times and baselines  as in Ref.~\cite{KAGRA:2021rmt}, and it reads
\begin{equation}
    \bm{\hat{\mathcal{P}}}_f = \bm{\Gamma}_f^{-1}\, \cdot\, \bm{X}_f\,, \quad \quad  \hat{\mathcal{P}}_\mu (f) = \sum_{\mu'}\,(\Gamma^{-1})_{\mu \mu'}(f)\,X_{\mu'}(f)\,.
\end{equation}
Here, $\bm{X}_f$ is the narrowband \textit{dirty map} and $\bm{ \Gamma}_f$ is the narrowband \textit{Fisher information matrix}, defined as follows \citep{Ain:2018zvo,KAGRA:2021rmt}:
\begin{equation}
\begin{aligned}
   \label{eq:X_Gamma_def}
   \bm{X}_f&\equiv \bm{\gamma}_f^\dagger \cdot \bm{N}_f^{-1} \cdot \bm{C}_f,\quad \quad X_\mu(f)  \propto \sum_{It;\,1,2 \in I} \frac{\gamma^{I*}_\mu(t;f)\,C^I(t;f)}{P_{{1}}(t;f)\,P_{{2}}(t;f)}\,,\\
   \bm{\Gamma}_f&\equiv \bm{\gamma}_f^\dagger \cdot \bm{N}_f^{-1} \cdot \bm{\gamma}_f,\quad \quad \Gamma_{\mu\mu'}(f)\propto\sum_{It;\,1,2 \in I}\frac{\gamma^{I*}_\mu(t;f)\,\gamma^{I}_{\mu'}(t;f)}{P_{{1}}(t;f)\,P_{{2}}(t;f)}\,,
\end{aligned}
\end{equation}
where $\bm{N}_f$ denotes the covariance matrix for the \ac{CSD} and $P_{1,2}(t;f)$ are the noise one-sided \acp{PSD}. The narrowband skymaps in the above equations are the foundation of all the anisotropic analyses in this paper.
\end{widetext}

\subsection{All-sky All-frequency radiometer search}\label{sec:ASAF_methods}
The \ac{ASAF} radiometer search targets point-like, persistent, narrowband \ac{GW} sources by scanning the frequency band and the sky using a {\tt HEALPix} grid ($N_\mathrm{side} = 16$)\footnote{This results in a sky grid with $12\, N_{\rm side}^{2} = 3072$ pixels, each covering $13.4\, {\rm deg^{2}}$.}~\cite{Gorski:2004by} and a frequency bin of $1/32$ Hz. Such sources can exhibit two types of waveforms in the time domain: (i) \ac{CGW} signal from a single \ac{GW} source, and (ii) a narrowband \ac{GWB} arising from multiple unresolved sources along a given line of sight, resulting in an incoherent signal. Being an unmodeled search, the radiometer search is robust against the signal model and serves as an alternative for detecting \ac{GW} signatures from poorly known sources, e.g., neutron stars having frequent glitches and/or accretion from a binary companion~\cite{Lasky:2015uia} or unknown \ac{CGW} sources \cite{Riles:2022wwz}.

We compute the point estimate, which serves as an estimator for the signal power and its associated uncertainty due to random noise. This is achieved by utilizing narrowband dirty maps and the Fisher information matrix shown in \eqref{eq:X_Gamma_def}, both constructed in the pixel basis, as described by the equations below
\begin{equation}\label{eq:ASAF_clean_sigma}
\begin{aligned}
    \hat{\mathcal{P}}_{\hat{n}}(f) &= \left[\Gamma_{\hat{n}\hat{n}}(f)\right]^{-1}\,X_{\hat{n}}(f)\,,\\ \sigma^2_{\hat{n}} (f)&= \left[\Gamma_{\hat{n}\hat{n}}(f)\right]^{-1}\,,
\end{aligned}
\end{equation}
where there is no summation over $\hat{n}$. Owing to the baseline's blind spots, the Fisher information matrix in the pixel basis is highly ill-conditioned~\cite{Mitra:2007mc,PhysRevD.100.043541,Renzini:2018nee,Renzini:2019vmt,Agarwal:2021gvz,PhysRevD.107.122002}. To avoid introducing numerical noise due to inversion of insensitive modes and their dependence on regularization, we proceed without incorporating pixel correlations. As shown in Ref.~\cite{Agarwal:2021gvz}, this approximation is justified given the current detector sensitivity when constructing clean map estimators. 

\subsection{Targeted-narrowband radiometer search}
\label{sec:nbr_methods}
The radiometer algorithm can be used in a pixel-based approach to target specific sky locations. For this analysis, five astrophysically motivated locations have been selected: three of them---Scorpius X-1, the Galactic Center, and the supernova remnant SN 1987A---were also analyzed in previous directional stochastic searches~\cite{LIGOScientific:2016nwa, LIGOScientific:2019gaw, KAGRA:2021mth}, while two globular clusters, Terzan 5 and NGC 6397, are introduced as new targets. These sources span a range of astrophysical scenarios, including accreting neutron stars, dense stellar environments, and young supernova remnants, and are all plausible hosts of \ac{CGW} emission \cite{Riles:2022wwz}. A brief overview of each source and its relevance to \ac{GW} searches is provided in \appref{sec:NBR_appendix_source}.

Unlike the \ac{ASAF} search, targeting specific locations allows us to account for the specific signal spread caused by Earth’s Doppler modulation or the intrinsic characteristics of the source. To do so, we combine the information from nearby frequency bins by performing a running average throughout the entire search frequency range. The characteristic window of the average is defined as the $N$ neighboring bins around a fixed frequency bin required to recover all the power spread by the phase evolution of the source. The output of the bin combination is a new power spectrum with the same original frequency resolution $\delta f$, but now accounting for the frequency variations of the signal throughout the observation time. The details regarding how the values of $N$ can be determined are given in \appref{sec:NBR_appendix_bin}.

\subsection{Broadband radiometer search}
\label{sec:BBR_methods}
The \ac{BBR} search is designed to measure the \ac{GWB} energy density from point-like sources across the sky and serves as a crucial tool for identifying persistent sources when there is stochasticity in the phase evolution of the signal due to a large number of sources. This search still relies on the same methods as in the ASAF analysis, while further assuming the point-like \ac{GWB} sources are broadband in frequency with an angular power spectral density $\mathcal{P}(f,\, \hat{n})$. Throughout this work, we search for \acp{GWB} whose $\mathcal{P}(f,\, \hat{n})$ can be factorized in one frequency-dependent and one angular-dependent factor \cite{KAGRA:2021mth,sgwb_unified_treatment, Allen:1996gp}, namely:
\begin{equation}
    \label{eq:BBR_methods_factorizability_P_f_n}
    \mathcal{P}(f,\, \hat{n}) = \mathcal{P}_{\hat{n}_{0}}(f) \, \delta^{2}(\hat{n}-\hat{n}_{0}) = \bar{H}(f)\, \mathcal{P}(\hat{n})\, ,
\end{equation}
where $\bar{H}(f)$ is the spectral shape of the \ac{GWB}, normalized such that $\bar{H}(f_{\rm ref})=1$, with $f_{\rm ref} = 25$ Hz \cite{KAGRA:2021kbb,KAGRA:2021mth}. We model $\bar{H}(f)$ as a simple power law, characterized by a spectral index $\alpha$ that is fixed throughout the search, such that \cite{sgwb_unified_treatment}
\begin{align}
    \label{eq:BBR_methods_frequency_power_law}
    &\bar{H}(f) \equiv \bar{H}(f;\, \alpha, \, f_{\rm ref}) = \left(\frac{f}{f_{\rm ref}}\right)^{\alpha -3}\, ,\\
    & \mathcal{P}(\hat{n}) \equiv \mathcal{P}(\hat{n}; \,\alpha,\, \,f_{\rm ref}) = \mathcal{P}_{\alpha,\,\hat{n}_{0}} (f_{\rm ref})\, \delta^{2}(\hat{n}-\hat{n}_{0})\, .
\end{align}
We consider three different \ac{GWB} power-law models: $\alpha = 0$, consistent with a cosmological GWB from slow-roll inflation or cosmic strings \cite{Caprini-Figueroa:2018mtu}; $\alpha = 2/3$, compatible with an astrophysical \ac{GWB} from \ac{CBC}s \cite{Regimbau:2011rp}; and $\alpha =3$, corresponding to a flat strain power spectrum~\cite{Allen:1997ad}. 

By following the same maximum-likelihood approach presented in \secref{sec:general_methods}, one can derive the expressions for the broadband dirty maps and Fisher matrix:
\begin{align}
    \label{eq:BBR_methods_maps}
    & X_{\hat{n}} = 
    \sum_{f} \bar{H}(f)\, X_{\hat{n}}(f)\, ,\\
    & \Gamma_{\hat{n} \hat{n}'} = \sum_{f} \bar{H}^{2}(f)\, \Gamma_{\hat{n} \hat{n}'}(f)\, ,
\end{align}
with $X_{\hat{n}}(f)$ and $\Gamma_{\hat{n} \hat{n}'}(f)$ from \eqref{eq:X_Gamma_def},
and the maximum-likelihood estimator 
for $\mathcal{P} (\hat{n})$
\begin{align}
    \label{eq:BBR_methods_weighted_estimators}
    \hat{\mathcal{P}}_{\hat{n}} = \frac{\sum_{f} \bar{H}(f)\sigma_{\hat{n}}^{-2}(f) \hat{\mathcal{P}}_{\hat{n}}(f)}{
    \sum_{f'} \bar{H}^{2}(f') \sigma_{\hat{n}}^{-2}(f')}\, .
\end{align}
We rescale this estimator to express it in units of the \ac{GW} energy flux spectrum at the reference frequency $f_{\rm ref}$, where $\mathcal{P}(f_{\rm ref}, \hat{n}) =  \mathcal{P}_{\alpha,\hat{n}}$ given $\bar{H}(f_{\rm ref})=1$, as
\begin{equation}
    \label{eq:BBR_methods_flux_estimator}
    \hat{\mathcal{F}}_{\alpha, \hat{n}}(f_{\rm ref}) = \frac{\pi c^{3}}{4G}\, f_{\rm ref}^{2} \hat{\mathcal{P}}_{\alpha, \hat{n}}\, .
\end{equation}

\subsection{Spherical harmonics search}
\label{sec:sph_methods}
This analysis is intended to search for an anisotropic distribution of spatially extended sources with a broadband spectrum, as opposed to the point sources targeted by the radiometer analyses discussed above.
While we follow the same factorization as given by \eqref{eq:BBR_methods_factorizability_P_f_n} of the \ac{BBR} search, we adopt the \ac{SPH} functions as a basis to characterize the anisotropies of extended sources~\cite{PhysRevD.80.122002}
\begin{align}
\calP(f, \hat{n})=\bar{H}(f)\sum_{\ell=0}^{\ell_\mathrm{max}}\sum_{m=-\ell}^\ell \calP_{\ell m}\,Y_{\ell m}(\hat{n})\, ,
\end{align}
where $ Y_{lm}(\hat{n}) $ is an \ac{SPH} function of the mode $(\ell, m)$ evaluated at the sky position $ \hat{n} $\, .
Since now we search for a \ac{GWB} signal with a broadband spectrum, we construct a \textit{broadband} version of the $\bm{\phat}$ estimator from its narrowband counterpart, following the derivation in \eqref{eq:BBR_methods_weighted_estimators}
but adapted to the \ac{SPH} basis. The $\bm{\phat}$ estimator is referred to as the \textit{clean map}, and the inversion of the Fisher matrix physically represents the deconvolution of the antenna pattern and detector noise.

To interpret the anisotropies of \acp{GWB} under the assumption of statistical isotropy, we
are more interested in the angular power of estimated $\phat_{\ell m}$, rather than their specific realization on the sky. Therefore, we introduce the estimator
of the angular power spectrum in the unit of $\mathrm{sr}^{-2}$
\begin{align}\label{eq:SPH_methods_curr_estimator}
\hat{C_\ell}=\left(\frac{2\pi^2f^3_\mathrm{ref}}{3H_0^2}\right)^2\frac{1}{2\ell+1}\sum_{m}\left[|\phat_{\ell m}|^2-(\Sigma_{\mathrm{R}})_{\ell m, \ell m}\right]\,,
\end{align} with its variance (under the weak-signal limit) given by:
\begin{align}\label{eq:SPH_methods_curr_estimator_sigma}
\mathrm{Var}[{\hat{C_\ell}}]\approx\left(\frac{2\pi^2f^3_\mathrm{ref}}{3H_0^2}\right)^4\frac{2}{(2\ell+1)^2}\sum_{m, m^\prime}\left|(\Sigma_{\mathrm{R}})_{\ell m, \ell m^\prime}\right|^2\,,
\end{align}
$\bm{\Sigma}_{\mathrm{R}}$ is the covariance matrix of the \textit{regularized} clean-map estimator, defined as:
\begin{align}\label{eq:SPH_methods_sigmas_curr_C_l_estimator}
    \bm{\Sigma}_{\mathrm{R}} = \bm{\Gamma}^{-1}_\mathrm{R}\cdot \bm{\Gamma}\cdot\bm{\Gamma}^{-1}_\mathrm{R}\,,
\end{align}
where $\bm{\Gamma}$ is the Fisher matrix of the dirty map on the \ac{SPH} basis and $\bm{\Gamma}_\mathrm{R}$ is its regularized version, whose details can be found in \appref{app:SPH_reg_fisher}.

Also, due to the angular resolution limit, we perform the analysis for \ac{SPH} modes only up to the angular scale given by $\lmax=3,4,16$ for $\alpha=0,2/3,3$,
respectively~\cite{LIGOScientific:2016nwa,LIGOScientific:2019gaw,KAGRA:2021mth}.
Technical details concerning the Fisher-matrix regularization as well as the angular resolution are discussed in \appsref{app:SPH_angres}{app:SPH_reg_fisher}. We eventually compare this $\hat{C}_\ell$ estimator to its theoretical predictions to characterize a potential signal or to place constraints on relevant theoretical models.

The estimator described above, which we refer to as the
\textit{auto}-$\hat{C}_\ell$ estimator, has been used in previous \ac{LVK}
analyses~\cite{LIGOScientific:2016nwa,LIGOScientific:2019gaw,KAGRA:2021mth}. However, Ref.~\cite{PhysRevD.109.103535} showed that in the
presence of an astrophysical \ac{GWB}, it is significantly biased by the
shot-noise-dominated nature of the signal, which arises from the discrete spatial or temporal realization of individual events. 
This consideration motivates us to also adopt the \textit{cross}-$\hat{C}_\ell$ estimator, defined as
\begin{equation}
\label{eq:SPH_methods_opt_estimator}
    \hat{C}_\ell^{\mathrm{(cross)}}=\left(\frac{2\pi^2f^3_\mathrm{ref}}{3H_0^2}\right)^2\frac{1}{2\ell+1}\frac{1}{n(n-1)}\sum_m\sum_{i\ne j}\phat^{(i)}_{\ell m}\,\phat^{(j)*}_{\ell m}\, ,
\end{equation}
with its variance (under the weak-signal limit\footnote{While the weak-signal limit mentioned in \secref{sec:general_methods} assumes the detector noise to dominate over the \ac{GWB} signal and astrophysical shot noise in the \textit{individual} time-frequency components, \eqsref{eq:SPH_methods_curr_estimator_sigma}{eq:SPH_methods_opt_estimator_sigma} involve the whole dataset and subsets, respectively, after marginalizing across times and frequencies.
}
)
\begin{align}
\label{eq:SPH_methods_opt_estimator_sigma}
\begin{split}
\mathrm{Var}[\hat{C_\ell}]&\approx\left(\frac{2\pi^2f^3_\mathrm{ref}}{3H_0^2}\right)^4\frac{2}{(2\ell+1)^2}\frac{1}{n^2(n-1)^2}\\
&\qquad\times\sum_{m, m^\prime}\sum_{i\neq j}(\Sigma_{\mathrm{R}}^{(i)})_{\ell m, \ell m^\prime}(\Sigma_{\mathrm{R}}^{(j)})_{\ell m^\prime, \ell m}\, ,
\end{split}
\end{align}
where $n$ distinct clean maps are involved and $\bm{\Sigma}_{\mathrm{R}}^{(i)}$ represents a covariance matrix of regularized clean maps derived from $i$-th dataset. See \appref{app:SPH_cross_Cell} for how we divide the whole dataset into subsets in the optimal way.

To either claim a detection of an anisotropic \ac{GWB} or place \acp{UL} on the angular power spectrum, given the observed $\phat_{\ell m}$ estimator, we evaluate its statistical significance. Specifically, we compute $p$-values for the real and imaginary parts of the ${\hat{\boldsymbol{\mathcal{ P}}}}$ estimator, respectively, in each \ac{SPH} mode using its expected \ac{PDF} based on Monte Carlo samples with the detailed procedure discussed in \appref{app:SPH_ul_pvalue}.
We generate 50,000 such $\bm{\phat}^\mathrm{sim}$ samples for the whole dataset and compute a $p$-value for each real and imaginary part of each \ac{SPH} mode. As a significance indicator, we adopt a 5\% $p$-value threshold, which we refer to as a \textit{local} threshold. Additionally, due to the presence of multiple observations across all the \ac{SPH} modes, we account for the trials factor in a conservative way by dividing each local threshold by the number of \ac{SPH} modes, $(\lmax+1)^2$, to obtain the \textit{global} $p$-value threshold.

\section{Results}
\label{sec:Results}
To perform all the four analyses described above, we analyze data from \ac{O1} and \ac{O2} of the LIGO detectors \cite{Advanced_LIGO_LIGOScientific:2014pky, Abbott:2016xvh, Nuttall:2015dqa,LIGOScientific:2017tza,LSC:2018vzm,LIGOScientific:2018kdd, Davis:2018yrz} located in Hanford (H) and Livingston (L); from \ac{O3} of LIGO \cite{aLIGO:2020wna,Tse:2019wcy,LIGO:2021ppb, Sun:2020wke} and Virgo \cite{advanced_Virgo_VIRGO:2014yos,Virgo:2019juy, Virgo:2022ysc,Virgo:2022kwz} (V); and from \ac{O4a} of
LIGO \cite{LIGO:2024kkz, Capote:2024rmo, LIGOO4Detector:2023wmz, membersoftheLIGOScientific:2024elc}.

These datasets are preprocessed following the procedure described in \cite{O4a_ISO}. Time-domain cuts are applied by removing segments affected by non-Gaussian features, hardware injections, and known instrumental artifacts; applying non-stationarity cuts; and gating loud glitches. These cuts are identical to those used in \cite{O4a_ISO}. In addition, frequency-domain cuts are applied to remove bins identified through coherence studies as contaminated by instrumental artifacts. Details of the observing run, including the time- and frequency-domain cuts and the effective dataset, are provided in \tabref{tab:DATA} of \appref{app:Data}.

We compute the \ac{CSD} by combining the \acp{SFT} of 192-second segments from all detector pairs, using a coarse-grained frequency resolution of $\Delta f = 1/32$ Hz and including the corresponding variances.

To reduce computational and storage costs in \ac{GWB} searches, we fold the 192-second segments across a sidereal day (23h 56m 4s), leveraging the temporal symmetry of the \ac{ORF} due to Earth’s rotation \cite{Ain:2015lea,Thrane:2015aua}. This preserves all necessary information while enabling efficient maximum-likelihood analysis over the full run.

We generate folded datasets for all observing runs~\cite{O3_FoldedData} up to and including \ac{O4a}, and perform all analyses using the {\pystoch} package \cite{Ain:2018zvo,Suresh:2020khz}, which now provides a unified framework supporting both pixel- and SPH-based analyses from \ac{O4a} onward.

\subsection{All-sky All-frequency radiometer search}
\begin{table*}[t]
\begin{tabular}{c@{\hspace{12pt}}|c@{\hspace{12pt}}|c@{\hspace{12pt}}|c@{\hspace{12pt}}|c@{\hspace{12pt}}|c}
\toprule
Run &~~Vetoed-Frequency~~& Range&~~Median~~&~~Median~~&~~Median (Mean) Ratio~~\\
 &~~Fraction (\%)~~&~~($\times 10^{-26}$)~~&~~($\times 10^{-26}$)~~&~~Common $f$ ($\times 10^{-26}$)~~&~~$\frac{\mathrm{Run}}{\mathrm{O1-O3}}$  \\
\midrule
\ac{O4a}  & 7.6 &3.46 - 223.12& 9.00 & 9.29     & 0.89 (0.72)  \\
O1-O3  & 1.2 &3.39 - 963.23 & 10.04 & 10.39     & 1  (1) \\
O1-\ac{O4a}  & 0.4 &2.94 - 845.68& 8.41 & 8.24     & 0.79 (0.67)   \\
\bottomrule
\end{tabular}
\caption{We present Bayesian sensitivity estimate on the effective strain amplitude $h_{\rm eff}$ for three datasets: O4a, O1–O3, and the combined O1–O4a. 
When comparing the median sensitivity of \ac{O4a} to \ac{O1}-\ac{O4a}, an improvement is observed. To account for differences in the vetoed frequency bins across observing runs, we also compare median sensitivity over the common frequency range.
}
\label{tab:ASAF_UL}
\end{table*}

\begin{figure*}
    \centering
    \begin{minipage}{0.48\textwidth}
    \includegraphics[scale=0.5]{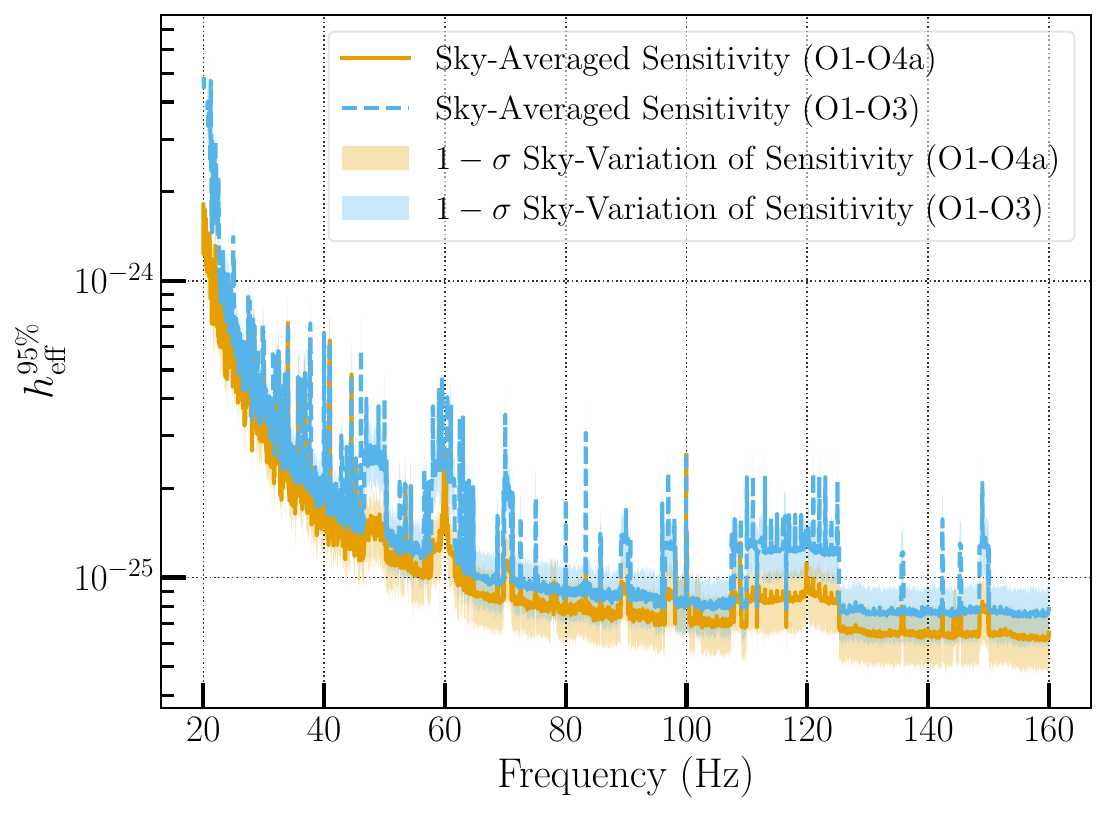}
    \end{minipage}   ~~~
   \begin{minipage}{0.48\textwidth}
        \includegraphics[scale=0.25]{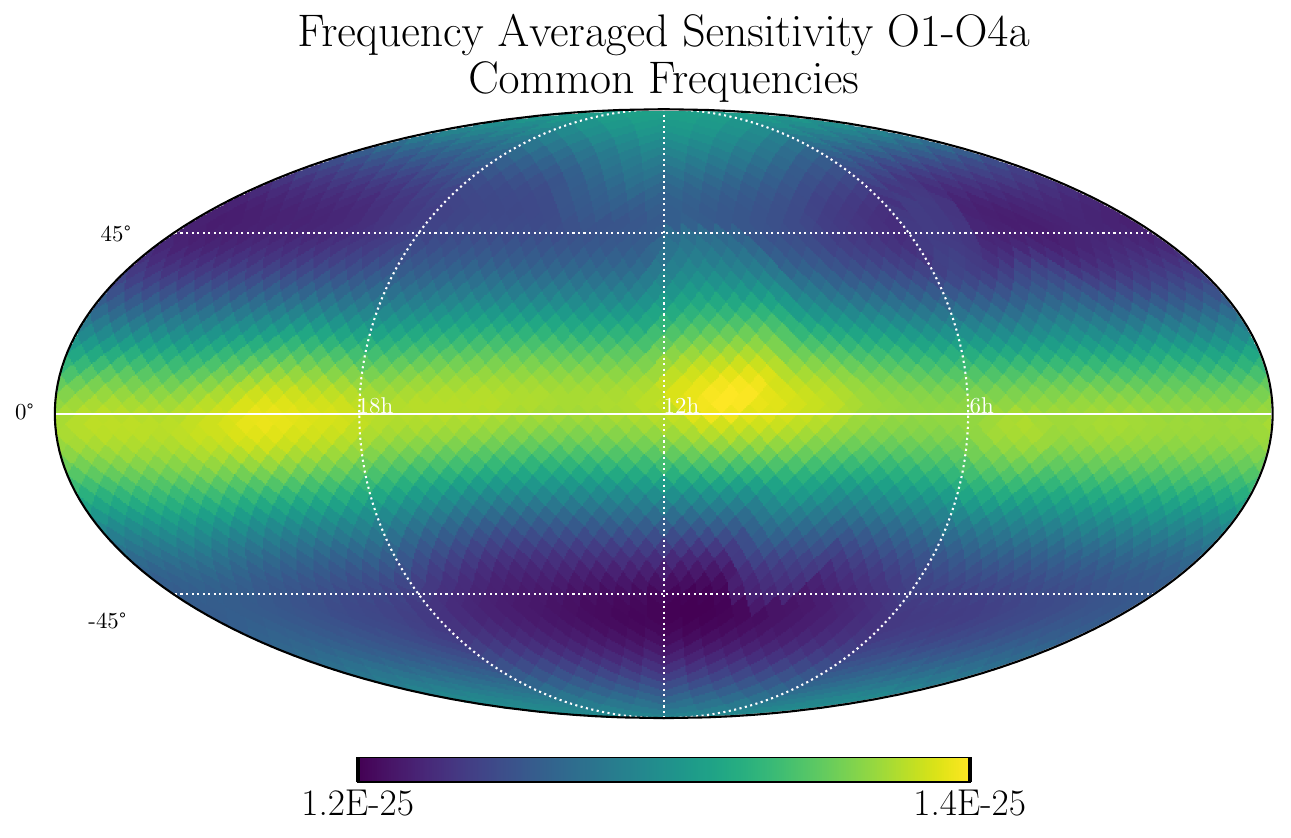}\\
   \includegraphics[scale=0.25]{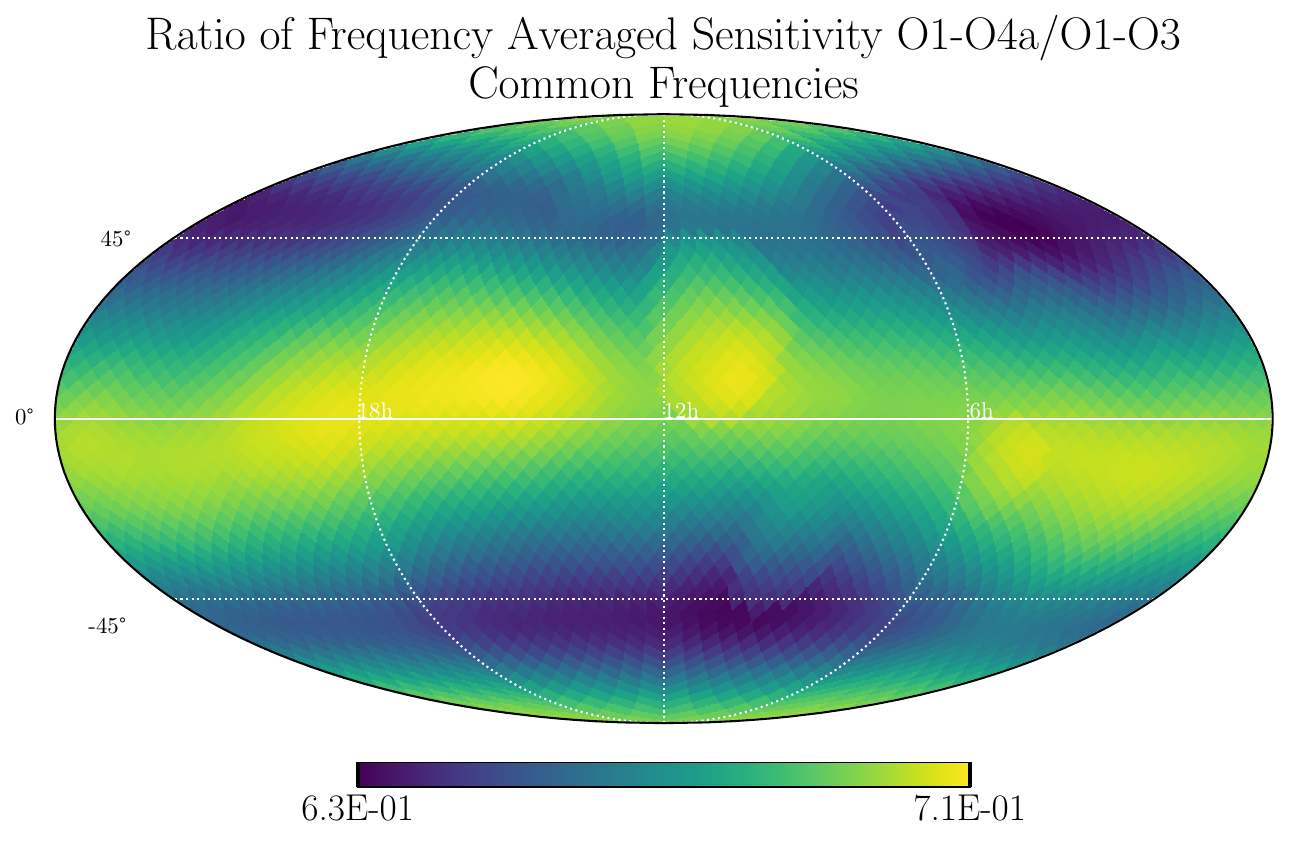}
   \end{minipage}
    \caption{Bayesian sensitivity estimate to Effective Strain Amplitude $h_{\rm eff}$:
   Left panel: Sky-averaged sensitivity from the \ac{O1}-\ac{O3} and \ac{O1}-\ac{O4a} datasets are shown as blue and yellow solid lines, respectively. The shaded regions represent 1-sigma variations in the estimates across the sky. The overall shape of the curves reflects the detectors' sensitivity profiles. Gaps indicate vetoed frequency bins, while narrow vertical peaks correspond to instrumental spectral lines that were not vetoed, as their impact on the analysis was minimal. 
    Top Right Panel: Skymap of frequency-averaged sensitivity using O1–O4a data. The observed pattern reflects the typical sky sensitivity of an HL-dominated network. As expected, the sensitivity is worse near the poles and equator, with improved sensitivity in the mid-declination regions. Bottom Right Panel: Skymap showing the ratio of frequency-averaged sensitivity from \ac{O1}-\ac{O4a} to that from \ac{O1}-\ac{O3}.     }
    \label{fig:ASAF_ULplot}
\end{figure*}

To identify a potential \ac{GW} signature in the data and set constraints on source parameters in the absence of a detection, we use the estimators constructed in \secref{sec:ASAF_methods} following the method outlined in Ref.~\cite{KAGRA:2021rmt} and detailed in \appref{app:ASAF_extra}. 

We use a detection statistic, the \ac{SNR},  defined as
\begin{equation}\label{eq:ASAF_SNR}
    \hat{\rho}_{\hat{n}}(f)\equiv\frac{\hat{\mathcal{P}}_{\hat{n}}(f)}{\sigma_{\hat{n}} (f)}\,,
\end{equation} 
to evaluate the significance of the data. The distribution of this statistic, obtained from \ac{O4a} data using both the random (unphysical) timeshift method~\cite{KAGRA:2021rmt} and the zero-lag data (i.e., data without timeshift), is shown in Appendix~\ref{app:ASAF_extra_sig}. The zero-lag data is mostly consistent with the random-timeshifted data within 2-sigma Poisson errors. With a global $p$-value threshold of 5\%, no evidence is found for a persistent narrowband \ac{GW} source. A total of 505 sub-threshold candidates identified for follow-up, and details are provided in Appendix~\ref{app:ASAF_extra_sig}.

Next, assuming that the \ac{GW} signal remains in the same frequency bin and sky direction, we combine data from the \ac{O1}–\ac{O4a} observing runs (including multiple baselines for \ac{O3}). This assumption sets an upper bound on the frequency drift of a \ac{CGW} signal:
\begin{equation}
   \frac{0.03125 \,\mathrm{Hz}}{8.3\,\mathrm{years}}=1.2\times10^{-10} \,\mathrm{Hz\,s}^{-1}\,.
\end{equation}
We note that incorporating data from multiple datasets reduced the notch fraction from 11.4\% in \ac{O4a} to 7.9\% in \ac{O1}–\ac{O4a}. The zero-lag data remains consistent with the null hypothesis. Additionally, we investigate the top three \ac{SNR} candidates in zero-lag with \ac{SNR}$>$5.5. We find that these candidates are associated with data quality issues at these frequencies, including hardware injections and calibration lines~\cite{LIGO:2024kkz}. Finally, a total of 562 sub-threshold follow-up candidates have been identified for the follow-up with methods tuned for searching for a \ac{CGW} (see Appendix~\ref{app:ASAF_extra_sig} for the details).

Having found the data to be consistent with Gaussian noise, we set Bayesian \acp{UL} on the effective strain amplitude, defined as
\begin{equation}
  h_{{\mathrm{eff}},\hat{n}}(f)\equiv \left[\mathcal{P}_{\hat{n}}(f)\, \Delta f\right]^{1/2}\,,
\end{equation}
with $95\%$ confidence. 
We note that the effective strain amplitude defined above is equal to the intrinsic strain amplitude $h_0$ for a circularly polarized \ac{CGW}, provided the signal remains confined within a single frequency bin~\cite{messenger2010understanding,Messenger:2015tga}]. In practice, this assumption is violated due to factors such as arbitrary polarization, Doppler shifts from the Earth’s orbital motion, and intrinsic source motion (e.g., proper motion or binary orbital motion), which introduce biases in the estimator. The Doppler shift from the Earth’s motion around the Sun can be approximated as $f_0 \times 2\times10^{-4}\cos\delta$, where $f_0$ is \ac{GW} frequency in the Solar System Barycentre frame and $\delta$ is the source declination. Consequently, above $\sim160\,\mathrm{Hz}$, the signal drifts beyond a single frequency bin. We therefore restrict our \ac{UL} estimates to the 20--160 Hz range, which we refer to as sensitivity estimates in further discussion. 

As shown in \tabref{tab:ASAF_UL}, the sensitivity estimate with \ac{O4a} and O1-O3 data across sky and frequencies lie in the ranges $(3.46-223.12)\times10^{-26}$ and $(3.39-963.23)\times10^{-26}$~\cite{KAGRA:2021rmt}, respectively. After combining all available data (i.e., O1-\ac{O4a}), the sensitivity lies in range $(2.94-845.68)\times10^{-26}$. 

To assess the improvement in the sensitivity estimate with the addition of the latest data, comparing only the minimum or maximum sensitivity across datasets is not a robust approach. This is primarily due to two reasons. First, specific frequency bins may be notched in one dataset but present in another, preventing a uniform comparison. For example, the notch fraction decreases from 7.6\% in \ac{O4a} to 0.4\% in the combined O1–O4a dataset (below 160 Hz). Second, the reported sensitivity incorporates both a point estimate and its associated uncertainty, both of which can fluctuate across the sky, introducing intrinsic statistical variability. We therefore compute the sensitivity improvement using the ratio of median sensitivity across these shared frequency bins. As shown in \tabref{tab:ASAF_UL}, the median sensitivity ratios for \ac{O4a} to O1–O3 and O1–O4a to O1–O3 are 0.89 and 0.79 respectively, indicating an overall improvement by a factor of 1.12 (1.26). 

To understand the variation of sensitivity across sky directions and frequencies, we consider a simplified model in which the sensitivity for each frequency-pixel pair is approximated as a combination of the point estimate and its associated uncertainty.
Given that the observed \ac{SNR} for frequency-pixel pairs is consistent with Gaussian noise with zero mean, the point-estimate $\hat{\mathcal{P}}_{\hat{n}}(f)$ can be assumed to be centered near zero. Consequently, the sky-averaged sensitivity can be approximated as (using Eqs.~\ref{eq:X_Gamma_def} and ~\ref{eq:ASAF_clean_sigma})
\begin{equation}
    {\rm Sensitivity}(f) \equiv \langle\sigma_{\hat{n}}(f)\rangle_{\hat{n}}^{1/2} \propto (P_1(f) P_2(f))^{1/4}\,,
\end{equation}
where $P_1(f)$ and $P_2(f)$ are noise PSD for detector 1 and 2 in baseline $I$. The sky-averaged sensitivity as a function of frequency is shown in the left panel of \figref{fig:ASAF_ULplot} for O1–O4a (blue) and O1–O3 (yellow). The trend follows the shape of the typical noise curve, i.e., $(P_1(f) P_2(f))^{1/4}$, with the shaded region indicating 1-sigma sky fluctuations.
Similarly, the frequency-averaged sensitivity for a given sky direction $\hat{n}$ is given by (using Eqs.~\ref{eq:X_Gamma_def} and ~\ref{eq:ASAF_clean_sigma})
\begin{equation}
    {\rm Sensitivity}_{\hat{n}} \equiv \langle\sigma_{\hat{n}}(f)\rangle_f^{1/2} \propto \left(\sum_t \Gamma_{\hat{n}}(t)\right)^{1/4}\,,
\end{equation}
where $\Gamma_{\hat{n}}(t)=\sum_A\, F^A_{1}(t,\hat{n}) \,F^A_{2}(t,\hat{n})$ is the combined antenna response~\cite{Mitra:2007mc}, and the summation is over observing time. This captures the directional variation in detector response over the full observing run. The corresponding skymap of frequency-averaged sensitivity for O1–O4a is shown in the top right panel of \figref{fig:ASAF_ULplot}. For the HL baseline-dominated data, the most sensitive sky regions lie between declinations of approximately $20^\circ$–$60^\circ$, with reduced sensitivity near the celestial poles and equator.

The bottom right panel of \figref{fig:ASAF_ULplot} presents the ratio of frequency-averaged sensitivity between O1–O4a and O1–O3 across the sky, which varies between 0.63 and 0.71.

 \subsubsection{Marginal Outlier at 92.8125 Hz} 

Signs of non-Gaussianity are observed in the negative \ac{SNR} tails of the zero-lag data, with the samples originating from the same frequency bin at 92.8125 Hz. While an actual astrophysical source is not expected to produce negative \ac{SNR} in our analysis, this candidate lies close to the \ac{SNR} threshold for a two-sided $p$-value. Therefore, a follow-up analysis was conducted, with details provided below.

In the \ac{O4a} data, we identified a frequency bin at 92.8125 Hz with extreme \ac{SNR} values: a minimum of $-6.12$ and a maximum of $5.2$ in the zero-lag dataset. The minimum \ac{SNR} lies in the negative tail and is just above the two-sided threshold of $-6.18$ corresponding to a 5\% $p$-value (see \ac{O4a} SNR histogram in Appendix~\ref{app:ASAF_extra}). While the positive \ac{SNR} is not significant enough to be classified as an outlier, the frequency-pixel pair was selected as a sub-threshold follow-up candidate. The associated \ac{SNR} skymap is given in Appendix~\ref{app:ASAF_extra}. In the random-timeshifted dataset, the \ac{SNR} range narrows to $-2$ to $2$, and the skymap structure disappears—consistent with behaviour expected in the case of a persistent signal. We investigate both tails to determine whether the excess power arises from detector noise or a \ac{GW} signal.

We note that, while most hardware injections were too weak to be confidently detected with our search method, the injection near 52.8 Hz was identified as a sub-threshold follow-up candidate, with an \ac{SNR} of 4.25—lower than that of the observed ``outlier". As shown in Appendix~\ref{app:ASAF_extra}, the \ac{SNR} in the outlier frequency bin steadily builds up in the zero-lag run and stands out compared to both neighbouring frequency bins and the random time-shifted dataset.

We also performed a simulation by injecting a \ac{GW} signal with \ac{SNR}$\sim$4.9 at the sky location corresponding to the observed maximum \ac{SNR}, convolved with the detector response, and added to numerous noise realizations. We find that, in approximately 4 out of $10^4$ trials, the negative \ac{SNR} exceeds the positive by a magnitude of 0.9, with the maximum \ac{SNR} exceeding 5.1, and the observed skymap is found to be a good match with the simulated map (see Appendix~\ref{app:ASAF_extra}).

When combining \ac{O4a} data with previous observing runs (\ac{O1}–\ac{O4a}), the \ac{SNR} decreases from $-6.12$ to $-4.3$ and from $5.2$ to $3.6$. For a persistent astrophysical source lasting $\sim$ 8 years, we would expect the \ac{SNR}—particularly in the positive direction—to increase if earlier data were sensitive and the signal remained in the same frequency bin. Otherwise, a decrease in \ac{SNR} upon combining runs may instead indicate a detector noise artifact that was absent in previous runs. We note that \ac{O3} HL data have sensitivity comparable to \ac{O4a} HL data, while other datasets were insufficiently sensitive. 

We do not see any excess auto-power localized in these frequency bins in the individual detectors; the nearest unknown excess power is observed in the L detector at around $92.7$ Hz (3 bins away). We have not identified any coherent instrumental witness channel that could account for the elevated \ac{SNR}. While some periods of elevated detector noise are present, there is no evidence of persistent or long-term non-stationary noise. Removing these periods from the analysed data does reduce the SNR, but the pattern in the sky persists.
The results remain inconclusive, and this frequency bin will be monitored in future observing runs to gather additional data.

\subsection{Targeted-narrowband radiometer search}
\label{sec:nbr_results}
\begin{figure*}
    \centering
    \includegraphics[width = 0.45\textwidth]{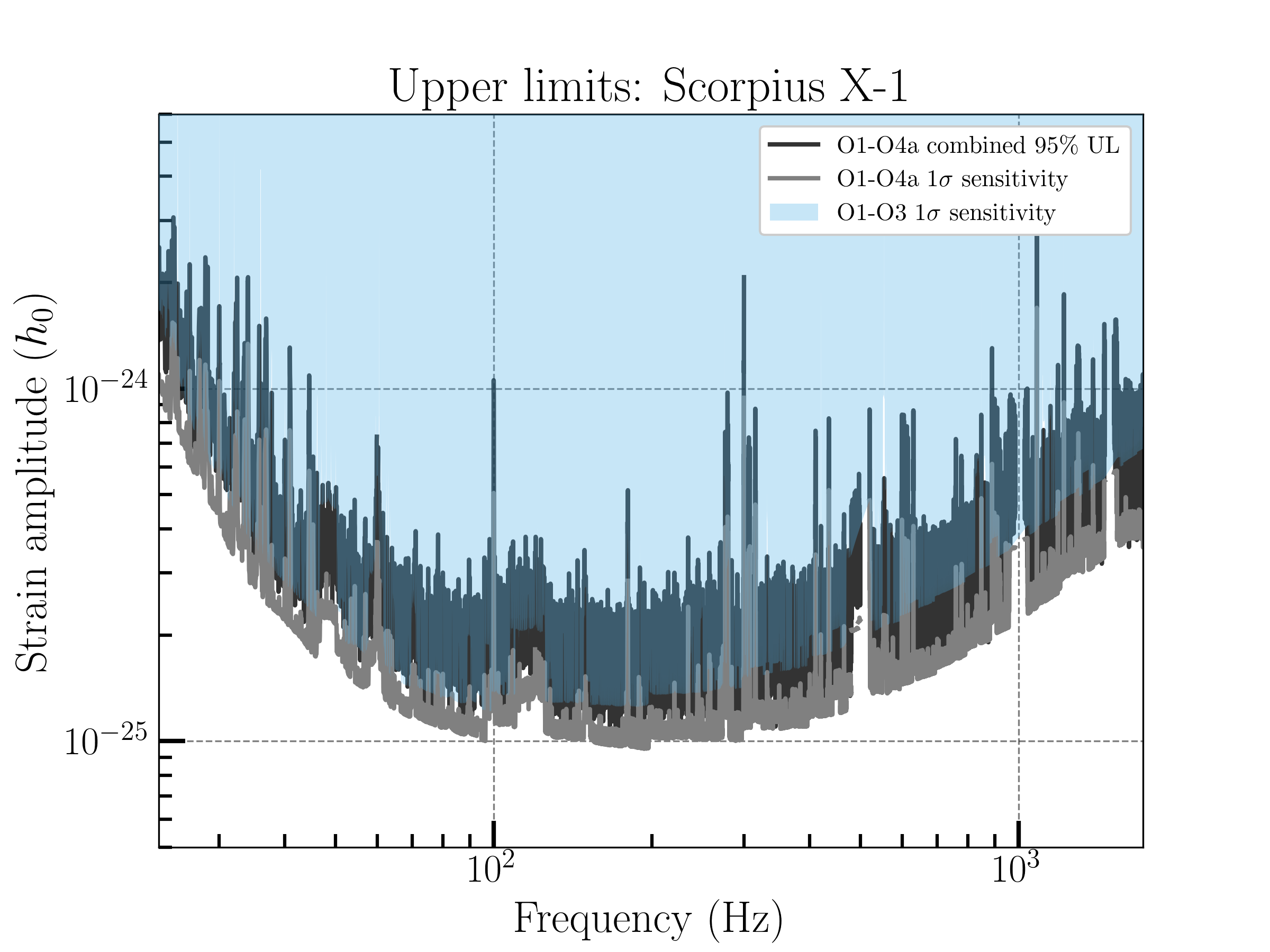}
    \includegraphics[width = 0.45\textwidth]{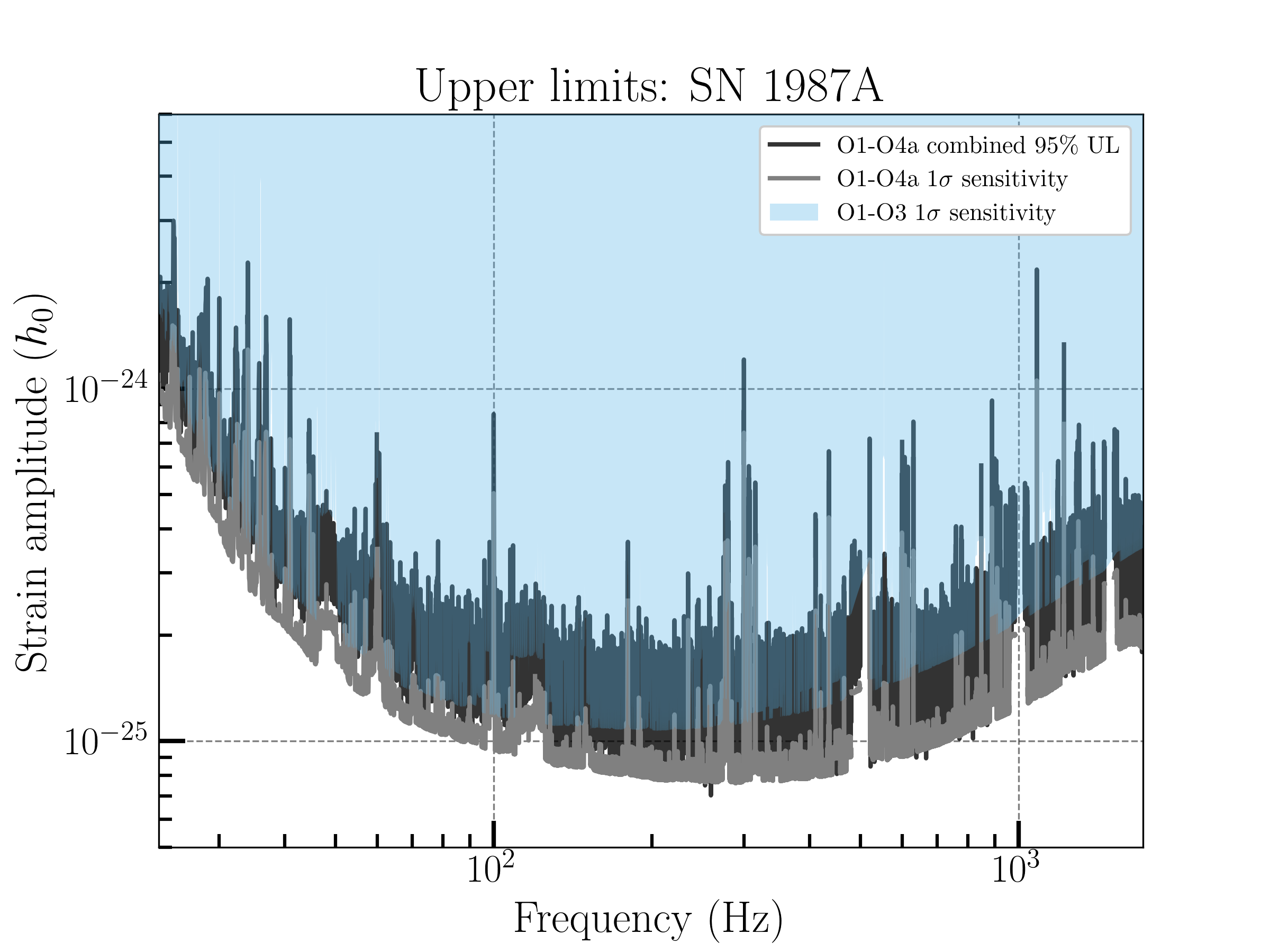}\\
    \includegraphics[width = 0.45\textwidth]{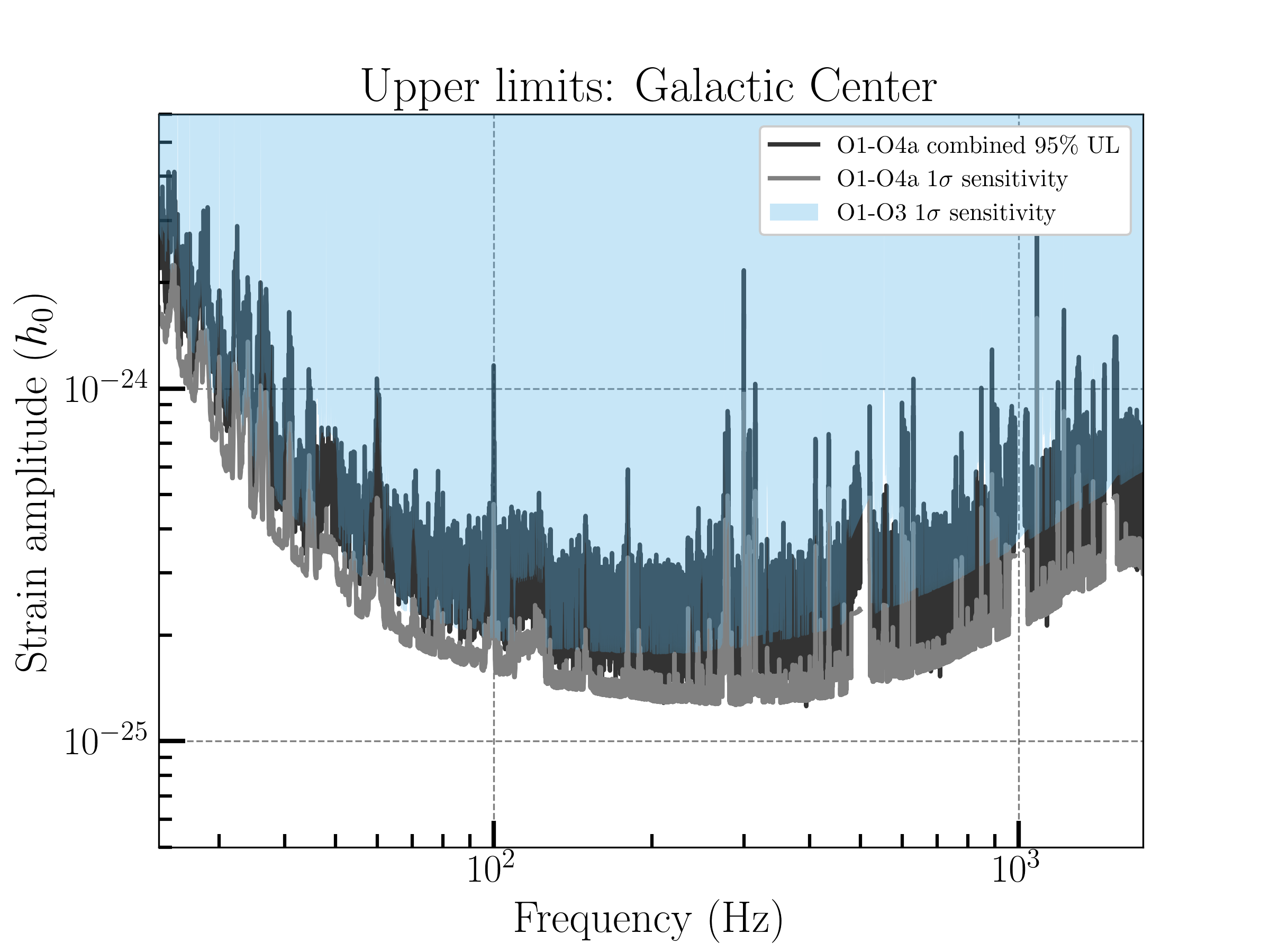}
    \includegraphics[width = 0.45\textwidth]{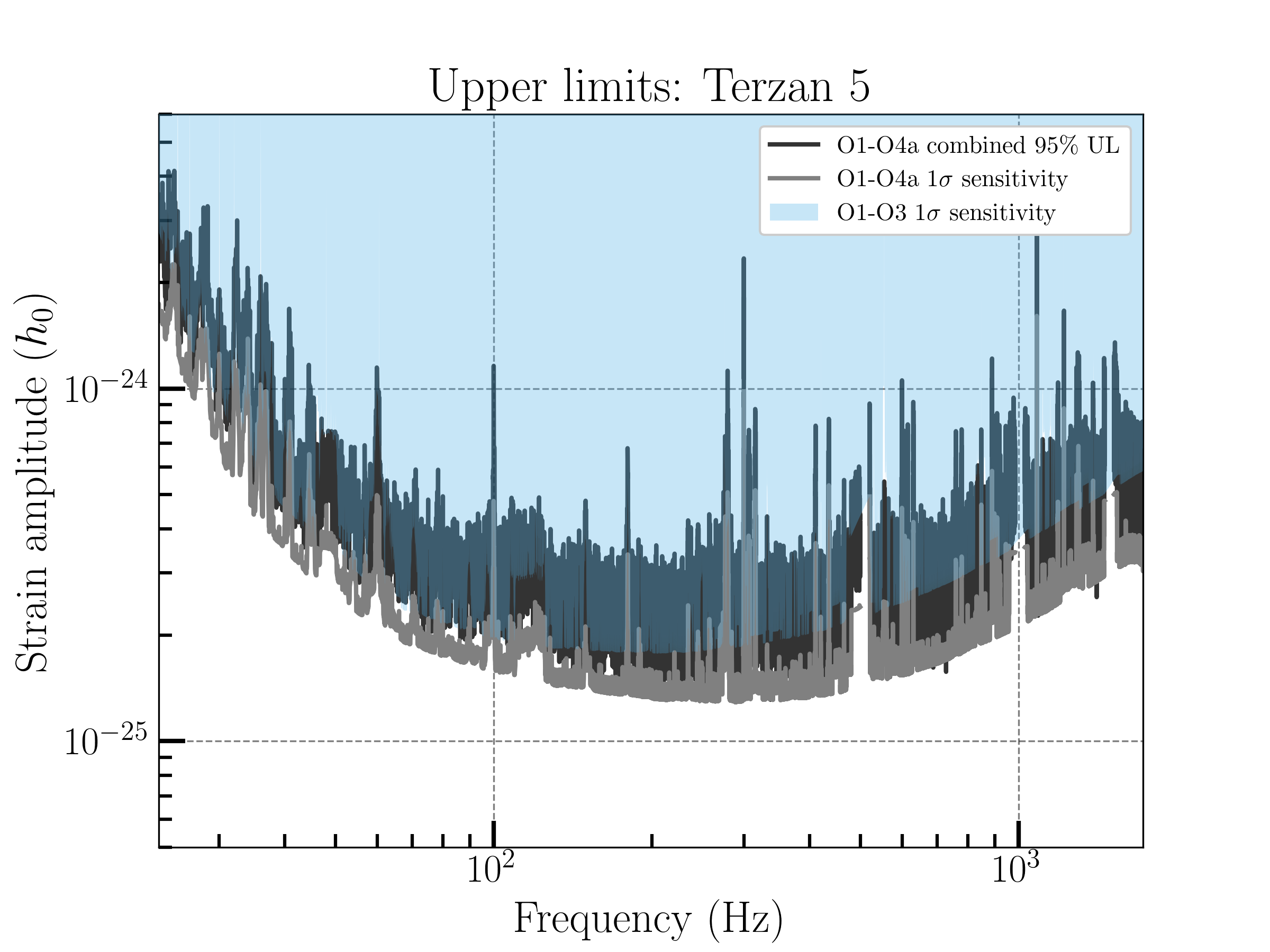}\\
    \includegraphics[width = 0.45\textwidth]{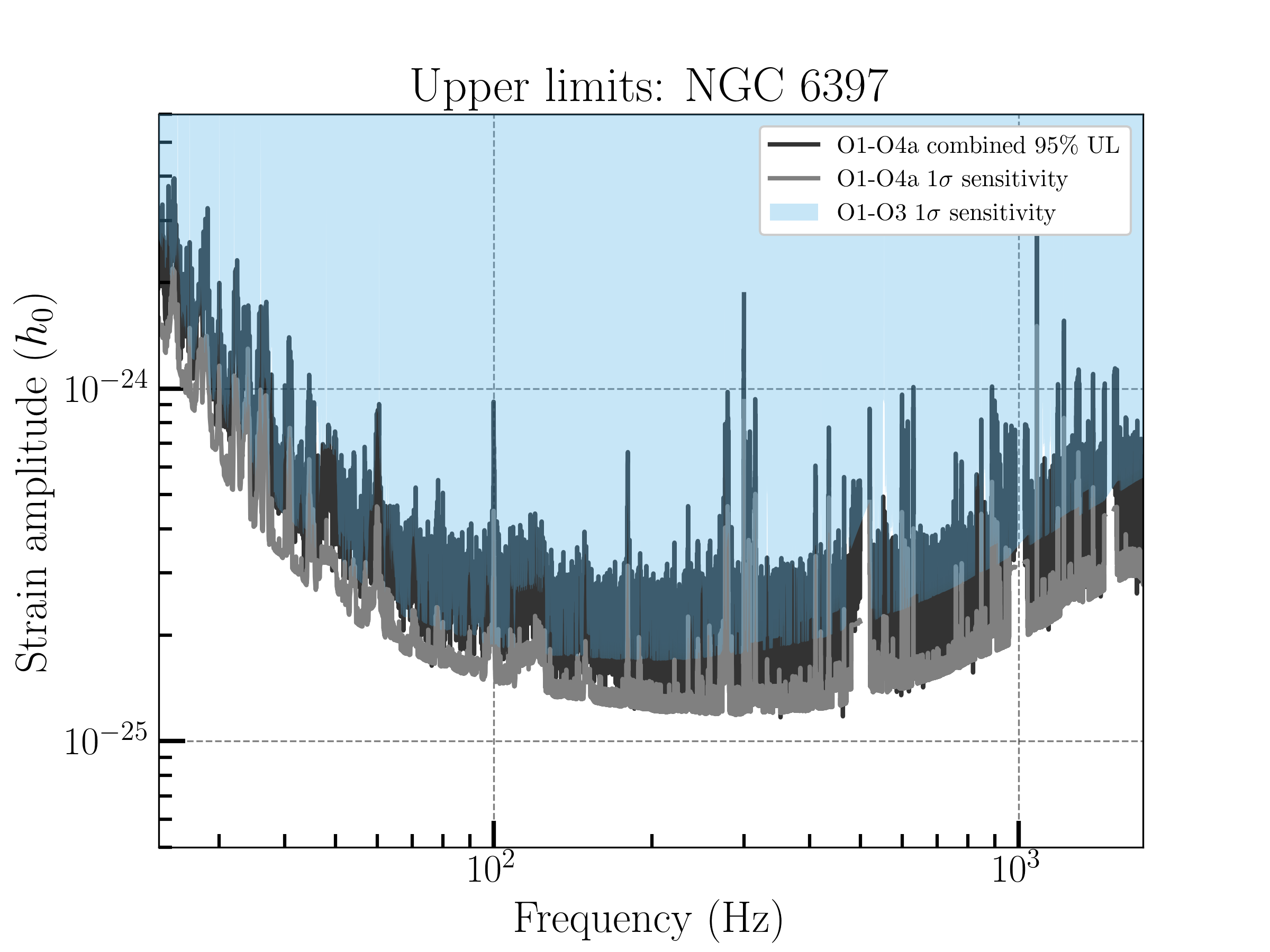}
    \caption{Plots showing the \acp{UL} for the five targets of the \ac{O1}-\ac{O4a} \ac{NBR} search. In each plot, the black solid line shows the Bayesian \acp{UL} set at 95\% confidence level, while the gray line is the $1\sigma$ sensitivity estimation in the hypothesis of no signal. The shaded blue area shows the $1\sigma$ sensitivity estimation with data up to the \ac{O3} run, highlighting the sensitivity improvement granted by adding the \ac{O4a} run to the analyses.}
    \label{fig:nbrcombul}
\end{figure*}

For all five directions, we computed the
\ac{SNR} by combining the appropriately sized frequency bins across the detectors. 
The best \acp{SNR} after bin combination are shown in \tabref{tab:nbrcombresults}, together with their $p$-values. 
The $p$-values are calculated from the maximum \ac{SNR} distribution obtained by simulating many realizations of strain power (details are shown in  \appref{sec:NBR_appendix_significance}).

Since we do not find any evidence for narrowband \acp{GW}, we place $95\%$ confidence \acp{UL} on the \ac{GW} strain spectrum from the five selected targets, shown in \figref{fig:nbrcombul} together with the $1\sigma$ sensitivity.
The results indicate a median improvement by a factor of 1.6–1.7 compared to previous results from \ac{O1} to \ac{O3}, with strain amplitudes ranging from $\sim 1.1 \times 10^{-25}$ to $6.5 \times 10^{-24}$ depending on the target.

Since the \ac{O4a} data have the best sensitivity so far, it is worth comparing the contribution that the \ac{O4a} dataset is giving to improve the reach of our searches. At higher frequencies, \ac{O4a} alone can provide better sensitivity than the combined O1 to O3 data. It also helps to improve the previous \acp{UL} in all the frequency ranges for all selected targets. The combination with future stages of the run promises to significantly increase the reach of our searches.

It is also useful to compare the results obtained with \ac{CGW} searches. \ac{CGW} searches use techniques that, for sources with at least a partial knowledge of the parameters, grant a much longer coherence time and hence better sensitivity than the targeted radiometer (for example, Ref.~\cite{LIGOScientific:2021quq} for the previous run and ~\cite{LIGOScientific:2025kei} for \ac{O4a}).  

Taking into account the several semi-coherent methods used for the \ac{O3} all-sky search for isolated pulsars~\cite{KAGRA:2022dwb}, the sensitivity estimation over the whole sky is comparable to the \acp{UL} shown in this section. However, more precise \ac{UL} estimation, like the one shown in the \ac{O3a} all-sky isolated \ac{CGW} search \cite{KAGRA:2021una}, can lead to significantly lower limits than the best ones obtained with SN 1987A of $h_0 \simeq 10^{-25}$. As a reference for the most recent SN 1987A dedicated search, see also the 90\% confidence \ac{UL} on \ac{O3} data in \cite{Owen:2023maw}.

A directed search to the Galactic Center in \ac{O3}, using a CW semi-coherent method, generally showed better sensitivity estimations than the targeted \ac{NBR} search~\cite{KAGRA:2022osp}.

Regarding Sco X-1, to assess the astrophysical significance of the signal strength probed in this search, a commonly used benchmark is the torque-balance level. In most \ac{CGW} searches of this target (see, for example, \cite{KAGRA:2022dqk, LIGOScientific:2022enz}, and the studies with updated ephemeris in \cite{Whelan:2023bha,Vargas:2023gvd}), an order-of-magnitude estimate of the torque-balance level is adopted as a reference point to evaluate whether the explored parameter space is astrophysically relevant. Given the absence of a detection in our search, we compare our \acp{UL} with this benchmark and find that they remain above the torque-balance level. 

These upper limits assume a generic polarization and marginalize over inclination and polarization angle (see \appref{sec:NBR_appendix_ul}). 
However, if we instead consider the circular polarization case—typically not quoted in our fully unmodeled analyses—our results indicate that the search has surpassed the torque-balance threshold under the hypothesis of circular polarization (more details are shown in \appref{sec:NBR_appendix_tb}).

\begin{table*}
\begin{tabular}{c@{\hspace{12pt}}c@{\hspace{12pt}}c@{\hspace{12pt}}c@{\hspace{12pt}}|c@{\hspace{12pt}}c}
\toprule
    {Direction} & {Max SNR}  & {Frequency band (Hz)} & {$p$-value (\%)} & {Best UL} ($\times 10^{-25}$)     & {Frequency band (Hz)}\\
\midrule
Scorpius X-1    & 4.2 & $1706.09375-1707.21875$  & 53.7  & 1.5 & $192.9375-193.0625$   \\
SN 1987A        & 4.1 & $1672.6875-1672.75$     & 59.0  & 1.1 & $246.75-246.8125$   \\
Galactic Center & 4.4 & $372.28125-372.90625$   & 22.5  & 1.9 & $227.4062-228.03125$   \\
Terzan5         & 4.5 & $821.46875-822.09375$   & 16.3   & 1.9 & $260.71875-261.34375$  \\
NGC6397         & 4.5 & $264.84375-265.40625$    & 18.2  & 1.7 & $251.40625-251.96875$   \\
\bottomrule
\end{tabular}
\caption{Results of the targeted-narrowband radiometer search on the O1-O4a combined datasets. We show the maximum SNR with its estimated $p$-value and frequency bin for each search direction. We also give the best 95\% confidence-level \ac{GW} strain \acp{UL}, with the corresponding frequency band, taken as the median of the most sensitive 1-Hz band.}
\label{tab:nbrcombresults}
\end{table*}
\subsection{Broadband radiometer search}
\label{sec:BBR_results}

\begin{table*}
    \begin{tabular}{c@{\hspace{12pt}}c@{\hspace{12pt}}c@{\hspace{12pt}}  c@{\hspace{12pt}}c@{\hspace{12pt}} |c@{\hspace{12pt}}c@{\hspace{12pt}}| c@{\hspace{12pt}}  c@{\hspace{12pt}} }
   \multicolumn{5}{c}{}& \multicolumn{2}{c}{{Max SNR} (\% $p$-value)} & \multicolumn{2}{c}{{UL ranges} $(10^{-8})$} $\mathrm{[erg\, cm^{-2}\, s^{-1}\, Hz^{-1}]}$\\
     \midrule
     &$\alpha$&  & $\Omega_{\rm gw}(f)$ & $\bar{H}(f)$ & HL(\ac{O4a}) &  O1+O2+O3+\ac{O4a} & O1+O2+O3+\ac{O4a} & O1+O2+O3 (HLV) \\
     \midrule 
     &0&			& constant 			&  $\propto f^{-3}$        &  1.7 (93) & 2.2 (72)   & 1.2 -- 5.5 & 1.7 -- 7.6   \\
     &$2/3$& 	& $\propto f^{2/3}$	&  $\propto f^{-7/3}$      &  1.8 (97) & 2.4 (76)  & 0.6 -- 3.0 & 0.85 --  4.1 \\
     &3&			& $\propto f^{3}$	&  constant       		 & 3.8 (20) & 3.8 (19) & 0.008 -- 0.092 & 0.013 -- 0.11  \\
     \bottomrule
   \end{tabular}
\caption{The maximum \ac{SNR} across all sky positions, its estimated $p$-value, and the range of the 95\% \acp{UL} on \ac{GW} energy flux $F_{\alpha, \hat{n}}^{95\%,\,\mathrm{UL}}(25\, {\rm Hz})$ $\mathrm{[erg\, cm^{-2}\, s^{-1}\, Hz^{-1}]}$ set by the \ac{BBR} search by combining the LIGO data from O1 to \ac{O4a} and the Virgo O3 data. The median improvement across the sky compared to limits from the O3 analysis is a factor of 1.4-1.7, depending on $\alpha$. O1+O2+O3 (HLV) \acp{UL} reported in the last column differ from the \acp{UL} in \cite{KAGRA:2021mth} due to the different $N_{\rm side}$ parameter we are employing here.}
\label{tab:BBR_O4a_results}
\end{table*}

\begin{figure*}
    \centering
    \includegraphics[width = \textwidth]{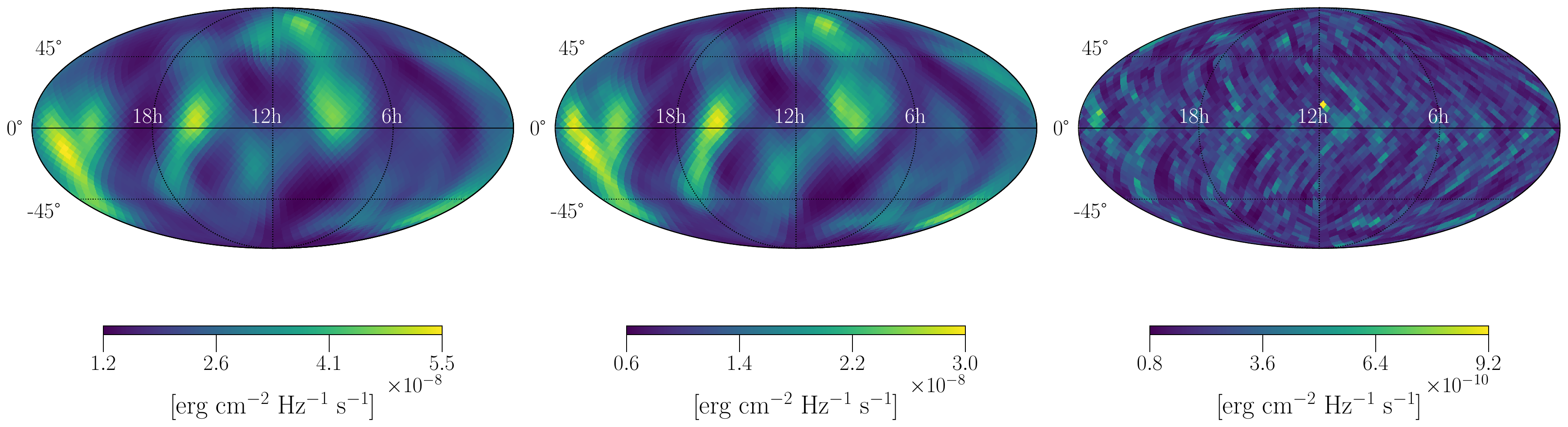}
    \caption{95\% confidence Bayesian \ac{UL} skymaps from a \ac{BBR} search for point-like sources. The maps are presented in the equatorial coordinate system and show UL skymaps of the \ac{GW} energy flux from the combination of all the data from O1 to \ac{O4a} LIGO observing runs and the Virgo O3 data. $\alpha =$ 0, 2/3, and 3 are represented from left to right.
    }
    \label{fig:BBR_O4a_ULs}
\end{figure*}

We present the results of the \ac{BBR} analysis for the \ac{O4a} dataset only and after combining them with the previously existing results from O1 to O3, which we also summarize in \tabref{tab:BBR_O4a_results}. We have performed such an analysis for three spectral indices, namely $\alpha= \{0,\,2/3,\,3\}$, whose final results are the skymaps shown in \figref{fig:BBR_O4a_ULs}. 
As in Ref.~\cite{KAGRA:2021mth} we pixelate the sky by employing the {\texttt{HEALPix}} scheme \cite{Gorski:2004by}, but instead of using $N_{\rm side} = 32$ as in ~\cite{KAGRA:2021mth}, here we use $N_{\rm side} = 16$.
We have opted for this choice as a compromise among the different angular resolutions of the baselines at the lower and upper ends of their most sensitive frequency bands, for different spectral indices $\alpha$ \cite{Mitra:2007mc, PhysRevD.106.023010}. 
We recall that we apply the same data quality prescriptions as detailed in \secref{sec:Results} and \appref{app:Data}.

Following the methods from \appref{sec:BBR_appendix}, we have first evaluated the \ac{SNR} maps for $\alpha =$ 0, 2/3, and 3, and then we have assessed the statistical significance of the maximum \ac{SNR} for each spectral index. 
We find that these \ac{SNR} values are consistent with the noise-only hypothesis, and we therefore proceed with the \ac{UL} calculation.
There is no evidence of outliers in the \ac{BBR} results, in contrast to the ASAF outlier in the 92.8125 Hz skymap. This is not inconsistent with the ASAF outlier, which is a narrowband feature and does not contribute significantly when integrating over the whole frequency band in the \ac{BBR} analysis \cite{O3_ASAF_data}.

\figref{fig:BBR_O4a_ULs} illustrates the skymaps showing the 95\% Bayesian \acp{UL} on the \ac{GW} energy flux $F_{\alpha, \hat{n}}(25\, {\rm Hz})$ in CGS units for $\alpha =$ 0, 2/3 and 3, obtained by combining the data from \ac{O1} to \ac{O4a}.
When comparing \figref{fig:BBR_O4a_ULs} to the results from the first three observing runs, we find that the median improvement in \acp{UL} across the sky is a factor of 1.4 for $\alpha = 0$, 1.4 for $\alpha = 2/3$, and 1.7 for $\alpha = 3$, respectively.
\subsection{Spherical harmonics search}
\label{sec:sph_results}
\begin{figure*}
\centering
\includegraphics[width=\textwidth]{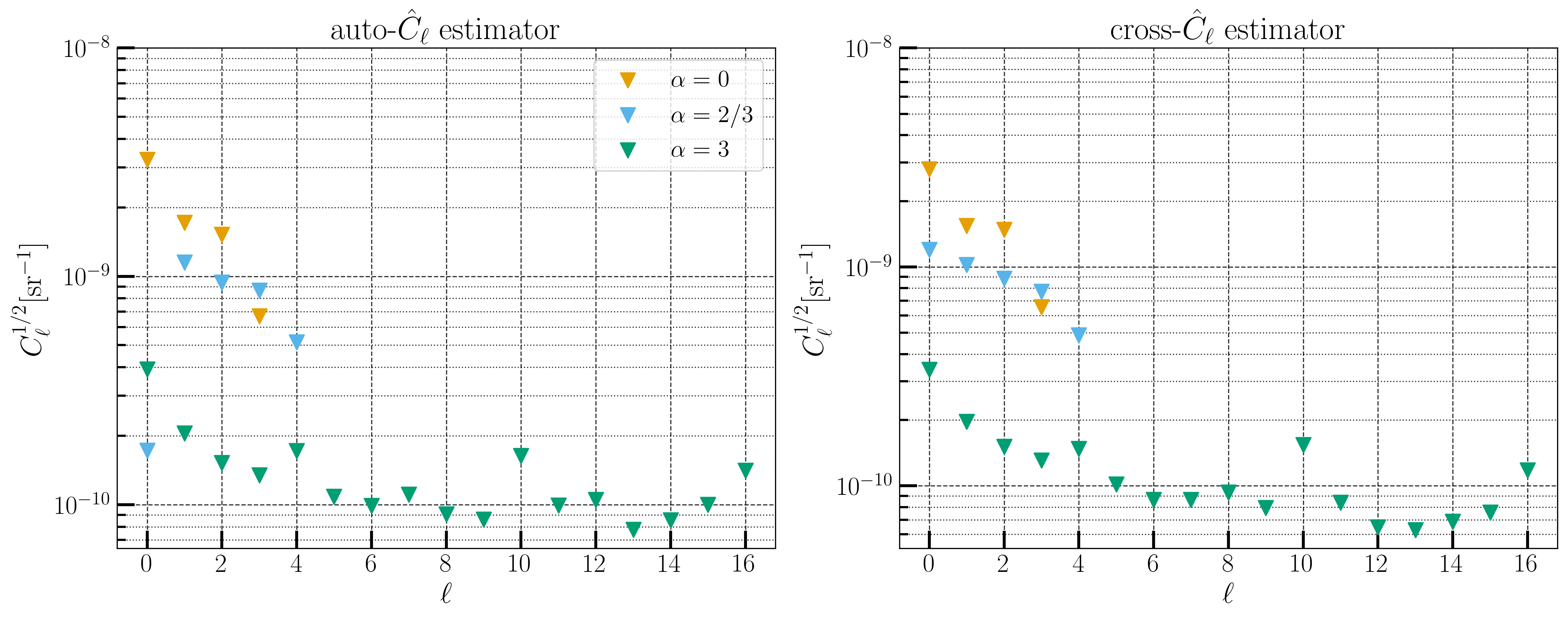}
\caption{$95\%$ \acp{UL} on the two $C_\ell$ estimators for different $\alpha$ using combined O1+O2+O3+O4a data.}
\label{fig:Cl_limits}
\end{figure*}

Following the procedure for significance assessment described in \secref{sec:sph_methods}, we compute the $p$-value for each real and imaginary part of the clean map estimator for the combined dataset across \ac{O1}-\ac{O4a}, and compare them to the local and global $p$-value thresholds, respectively. We find that all \ac{SPH} modes lie within the global $p$-value threshold across the three power-law spectrum models, indicating consistency with a Gaussian distribution. See \appref{app:SPH_ul_pvalue} for more details.
Given the absence of a detected \ac{GWB} signal, we compute \acp{UL} on
$C_\ell^{1/2}$ at each angular scale characterized by the $\ell$ value, for
different power-law frequency spectrum models.
\figref{fig:Cl_limits} shows the $C_\ell$ \acp{UL} derived from the two estimators for each power-law spectrum model. Also, the uncertainties of $\hat{C}_\ell$ measurement are improved by factors of 1.4-2.2 compared to the previous search, as discussed with more details in \appref{app:SPH_ul_pvalue}.

Note that the variance of the $\hat{C}_{\ell}$ estimator and its \acp{UL} depend on the definition of the $\hat{C}_\ell$ estimator or even the specific regularization we apply to the Fisher matrix. Therefore, one should not quantitatively compare the \acp{UL} reported here to those in Ref.~\cite{KAGRA:2021mth}, where the regularization was performed differently as compared to what is presented in this paper. The improvement factors mentioned above are based on the \acp{UL} recomputed for O1+O2+O3 data using the consistent regularization method, described in \appref{app:SPH_reg_fisher} for either definition of the $C_\ell$ estimators. For the same reason, these \acp{UL} cannot be compared between the two different estimators. Even though we apply the same regularization to both cases, the variance of the cross-${\hat{C}_\ell}$ estimator also depends on how the entire dataset is split into subsets since the inversion of a Fisher matrix is not a linear operation. Therefore, we treat these estimators as two different ways to present our search results.

Below, we consider the implications of our results for different astrophysical
models. For $\alpha=2/3$, the \ac{UL} found here for the corresponding
$\ell$ modes is $C_\ell^{1/2} < 1.2\times 10^{-9}\, \mathrm{sr}^{-1}$, whereas
several theoretical studies in the literature
\cite{Jenkins:2018uac,PhysRevLett.122.111101,PhysRevLett.120.231101} predict a
range of $C_{\ell}^{1/2} \sim (0.2-5)\times10^{-11} \, \mathrm{sr}^{-1}$ for
$1\leq \ell \leq 4$, assuming the normalized \ac{GW} energy density due to an
isotropic \ac{GWB} of compact binaries is $\Ogw\lesssim (2.2-6.7)\times
10^{-9}$~\cite{O4a_ISO}. Also, note that the shot noise term is expected to
be $(C_\ell^\mathrm{shot})^{1/2}\sim 10^{-10}\,\mathrm{sr}^{-1}$, dominating the
anticipated true astrophysical power spectrum~\cite{PhysRevD.100.063508}, and is
still below the obtained \acp{UL} by an order of magnitude. This is
consistent with the fact that the point estimates for the two types of the $\hat{C}_\ell$ estimators are not significantly different.

For $\alpha=0$, we find the \ac{UL} to be $C_\ell^{1/2} \leq (0.7-1.7)\times10^{-9}\,
\mathrm{sr}^{-1}$ for $1\leq\ell\leq3$, whereas the theoretical study on Nambu-Goto strings based on the model 3 in Ref.~\cite{PhysRevD.98.063509}, combined with the most up-to-date constraints on $G\mu$ using the isotropic component of the GWB \cite{O4a_CosmoImpli}, $G\mu\lesssim (2.7\sim 4.2)\times10^{-15}$, sets $C_1^{1/2}\lesssim 10^{-12}\,\mathrm{sr}^{-1}$.
For both choices of the power spectra ($\alpha = 0$ and $\alpha = 2/3$), we conclude that the predictions of the theoretical models are consistent with the search results presented here.

\section{Conclusions}
\label{sec:conclusion}
We do not find evidence for \ac{GW} signals in any of the four analyses using data from the first three observing runs of LIGO, Virgo, and \ac{O4a}. 
Hence, each analysis yielded the most stringent constraints to date from the directional search for persistent \acp{GW}.
For the all-sky all-frequency analysis, we observe a median (mean) improvement factor of 1.12 (1.38) in Bayesian sensitivity estimate to the effective strain amplitude when comparing \ac{O4a} to \ac{O1}–\ac{O3} in \tabref{tab:ASAF_UL}. Combining data from \ac{O1}–\ac{O4a} enhances this improvement by a factor of 1.13 relative to \ac{O4a} alone, and the fraction of notched frequency bins is reduced from 11.4\% to 7.9\%. For the targeted-narrowband radiometer analysis, \ac{O4a} contributes significantly to improve the upper limits of the entire frequency range (see \figref{fig:nbrcombul}), in particular at higher frequencies. In the broadband radiometer analysis, when comparing with the first three observing runs in \tabref{tab:BBR_O4a_results}, the median improvement across the sky in the upper limits on the \ac{GW} energy flux is 1.4, 1.4, and 1.7 for $\alpha = 0, 2/3, 3$, respectively. Lastly, the spherical harmonic analysis has introduced a cross-$C_\ell$ estimator to remove the potential shot noise and \ac{GWB} bias as opposed to the conventional auto-$C_\ell$ estimator. The uncertainty on the angular power spectrum, $C_\ell$, derived from each estimator has improved by a factor of 1.4-2.2 compared to the first three observing runs. Also, the upper limits for each estimator shown in \figref{fig:Cl_limits} are consistent with the predictions of the theoretical models, such as cosmic strings and kinematic dipole.

During \ac{O4a}, the Virgo detector was not in science mode and is therefore not included in our analysis. However, as noted in previous studies \cite{KAGRA:2021mth}, incorporating the Virgo detector into the network (even with its higher noise levels compared to the LIGO detectors) serves as a natural regularizer in extended source searches. This, in turn, enables us to resolve finer structures in the \ac{GW} skymaps.
As shown in \cite{Venikoudis:2024zmk}, current directional analyses are not affected by correlated noise, such as magnetic correlations between detectors. While these contributions are currently negligible, they are expected to become increasingly important as detector sensitivities improve. 

Improvements in detector sensitivity, longer observing runs, and larger network configuration will get us closer to the detection of potential local anisotropies or previously unknown point-like sources. It is also worth noting that the potential of all-sky all-frequency sources to uncover unknown narrowband signals represents an exciting direction for future investigations. A thorough investigation of the potential follow-up candidates from the all-sky all-frequency analyses, as presented in Ref.~\cite{Knee:2023toa}, could provide valuable insights into the nature and significance of the identified candidates.

Looking ahead, as the fourth observing run progresses, we will continue to incorporate additional data and update the findings reported in this paper.

\section*{Acknowledgements}
This material is based upon work supported by NSF's LIGO Laboratory, which is a
major facility fully funded by the National Science Foundation.
The authors also gratefully acknowledge the support of
the Science and Technology Facilities Council (STFC) of the
United Kingdom, the Max-Planck-Society (MPS), and the State of
Niedersachsen/Germany for support of the construction of Advanced LIGO 
and construction and operation of the GEO\,600 detector. 
Additional support for Advanced LIGO was provided by the Australian Research Council.
The authors gratefully acknowledge the Italian Istituto Nazionale di Fisica Nucleare (INFN),  
the French Centre National de la Recherche Scientifique (CNRS) and
the Netherlands Organization for Scientific Research (NWO)
for the construction and operation of the Virgo detector
and the creation and support  of the EGO consortium. 
The authors also gratefully acknowledge research support from these agencies as well as by 
the Council of Scientific and Industrial Research of India, 
the Department of Science and Technology, India,
the Science \& Engineering Research Board (SERB), India,
the Ministry of Human Resource Development, India,
the Spanish Agencia Estatal de Investigaci\'on (AEI),
the Spanish Ministerio de Ciencia, Innovaci\'on y Universidades,
the European Union NextGenerationEU/PRTR (PRTR-C17.I1),
the ICSC - CentroNazionale di Ricerca in High Performance Computing, Big Data
and Quantum Computing, funded by the European Union NextGenerationEU,
the Comunitat Auton\`oma de les Illes Balears through the Conselleria d'Educaci\'o i Universitats,
the Conselleria d'Innovaci\'o, Universitats, Ci\`encia i Societat Digital de la Generalitat Valenciana and
the CERCA Programme Generalitat de Catalunya, Spain,
the Polish National Agency for Academic Exchange,
the National Science Centre of Poland and the European Union - European Regional
Development Fund;
the Foundation for Polish Science (FNP),
the Polish Ministry of Science and Higher Education,
the Swiss National Science Foundation (SNSF),
the Russian Science Foundation,
the European Commission,
the European Social Funds (ESF),
the European Regional Development Funds (ERDF),
the Royal Society, 
the Scottish Funding Council, 
the Scottish Universities Physics Alliance, 
the Hungarian Scientific Research Fund (OTKA),
the French Lyon Institute of Origins (LIO),
the Belgian Fonds de la Recherche Scientifique (FRS-FNRS), 
Actions de Recherche Concert\'ees (ARC) and
Fonds Wetenschappelijk Onderzoek - Vlaanderen (FWO), Belgium,
the Paris \^{I}le-de-France Region, 
the National Research, Development and Innovation Office of Hungary (NKFIH), 
the National Research Foundation of Korea,
the Natural Sciences and Engineering Research Council of Canada (NSERC),
the Canadian Foundation for Innovation (CFI),
the Brazilian Ministry of Science, Technology, and Innovations,
the International Center for Theoretical Physics South American Institute for Fundamental Research (ICTP-SAIFR), 
the Research Grants Council of Hong Kong,
the National Natural Science Foundation of China (NSFC),
the Israel Science Foundation (ISF),
the US-Israel Binational Science Fund (BSF),
the Leverhulme Trust, 
the Research Corporation,
the National Science and Technology Council (NSTC), Taiwan,
the United States Department of Energy,
and
the Kavli Foundation.
The authors gratefully acknowledge the support of the NSF, STFC, INFN and CNRS for provision of computational resources.
This work was supported by MEXT,
the JSPS Leading-edge Research Infrastructure Program,
JSPS Grant-in-Aid for Specially Promoted Research 26000005,
JSPS Grant-in-Aid for Scientific Research on Innovative Areas 2402: 24103006,
24103005, and 2905: JP17H06358, JP17H06361 and JP17H06364,
JSPS Core-to-Core Program A.\ Advanced Research Networks,
JSPS Grants-in-Aid for Scientific Research (S) 17H06133 and 20H05639,
JSPS Grant-in-Aid for Transformative Research Areas (A) 20A203: JP20H05854,
the joint research program of the Institute for Cosmic Ray Research,
University of Tokyo,
the National Research Foundation (NRF),
the Computing Infrastructure Project of the Global Science experimental Data hub
Center (GSDC) at KISTI,
the Korea Astronomy and Space Science Institute (KASI),
the Ministry of Science and ICT (MSIT) in Korea,
Academia Sinica (AS),
the AS Grid Center (ASGC) and the National Science and Technology Council (NSTC)
in Taiwan under grants including the Science Vanguard Research Program,
the Advanced Technology Center (ATC) of NAOJ,
and the Mechanical Engineering Center of KEK.

\appendix
\section{Observing runs and dataset} \label{app:Data}

\tabref{tab:DATA} details each individual observing run of the detectors from \ac{O1} to \ac{O4a}, including their start and end times, the amount of data lost through time-domain cuts, and the effective data available for the \ac{GWB} directional searches. As part of the time-domain cuts, we exclude data contaminated by instrumental artifacts, hardware injections used for signal validation, and segments containing known \ac{GW} signals. We also apply a standard non-stationarity cut \cite{O4a_ISO} to eliminate segments that do not behave as Gaussian noise. The table also shows the fraction of frequency bins removed from the analysis. These removals follow the same frequency-domain cuts described earlier, based on coherence studies identifying contamination from instrumental artifacts. The specific bins removed may vary across analyses depending on sensitivity to narrow spectral features.

\begin{table*}[t]
    \begin{tabular}{c@{\hspace{5pt}}c@{\hspace{10pt}}c@{\hspace{2pt}}c|@{\hspace{8pt}}c@{\hspace{12pt}}c@{\hspace{8pt}}|c@{\hspace{2pt}}|c}
        \toprule
       \multirow{2}{*}{\shortstack{{Run} \\ \cite{LIGOScientific:2016nwa,KAGRA:2021rmt,LIGOScientific:2019gaw,KAGRA:2021mth}}} & 
        \multirow{2}{*}{\shortstack{{Start time} \\(UTC)}} &
        \multirow{2}{*}{\shortstack{{End time}\\ (UTC)}} & 
        {Detectors} & 
        \multirow{2}{*}{\shortstack{{Time} \\ {domain cut}\\ (\%)}} & 
        \multirow{2}{*}{\shortstack{{Effective data}\\ (days)}} & 
\multicolumn{2}{c}{\shortstack{Frequency notch fraction (\%)}} \\
& & & & & & \shortstack{\ac{ASAF} analysis} & \shortstack{Other analyses} \\
        & & & & & &  & \\
        \midrule
        O1   & 2015-09-18 15:00 & 2016-01-12 16:00 & H, L      & 35 & 29.85 & 25.5 & 21.1\\[2mm]
        O2  & 2016-11-30 16:00 & 2017-08-25 22:00 & H, L      & 16 & 99 & 19.6 & 15.3\\ [2mm]
        \multirow{3}{*}{O3}  & 
        \multirow{3}{*}{2019-04-01 15:00} & 
        \multirow{3}{*}{2020-03-27 17:00} & 
        \multirow{3}{*}{H, L, V}   & 
        \multirow{3}{*}{\shortstack{10.7 \tiny{(HL)} \\  14.3 \tiny{(HV)} \\ 14.7 \tiny{(LV)}}} & 
        \multirow{3}{*}{\shortstack{169 \tiny{(HL)} \\ 146 \tiny{(HV)} \\ 153 \tiny{(LV)}}} & \multirow{3}{*}{\shortstack{21 \tiny{(HL)} \\ 34.8 \tiny{(HV)} \\ 28.15 \tiny{(LV)}}} & \multirow{3}{*}{\shortstack{14.8 \tiny{(HL)} \\ 25.2 \tiny{(HV)} \\ 21.9 \tiny{(LV)}}}\\
                & & & & &\\
        & & & & &\\[2mm]

        O4a & 2023-05-24 15:00& 2024-01-16 16:00 & H, L      & 8.3 & 108.65 & 11.4 & 11.4\\

        \bottomrule
    \end{tabular}
    \caption{This table shows the individual observing runs, their start and end times, the detectors involved, the data quality cuts (both time and frequency domain) applied, and the effective data used in this analysis.}
    \label{tab:DATA}
\end{table*}

\section{ASAF radiometer search}\label{app:ASAF_extra}

\subsection{Significance}\label{app:ASAF_extra_sig}

We summarize the statistical framework used here to identify the \ac{GW} signal in the \ac{ASAF} search.

The null hypothesis assumes that the data contain only Gaussian noise, while the alternative hypothesis is that a \ac{GW} source is present in at least one frequency-pixel pair. The detection statistic is the \ac{SNR} (\eqref{eq:ASAF_SNR}), which under the null ideally follows a zero-mean Gaussian distribution.

Because real detector data include non-Gaussian features such as narrowband artifacts and glitches, the null \ac{SNR} distribution is obtained using the random time-shift (TS) method~\cite{KAGRA:2021rmt}. The procedure involves dividing the data from the entire observing run into multiple jobs (2493 in the case of O4a), each typically having a maximum duration of 5000 seconds. For each job, the data from one detector is held fixed while the data from the other detector is shifted by a random time delay, uniformly sampled between 1 and 2 seconds. This random time-shifting process effectively eliminates any coherent and persistent signals, ensuring that only the noise background remains in the data~\cite{KAGRA:2021rmt}.

After ensuring the Gaussianity of the null distribution, we proceed to test the zero-lag (ZL) data (without an unphysical time-shift) against the null hypothesis. We determine the observed highest \ac{SNR} across frequency-pixel pairs and compute the local $p$-value, $p_L$, assuming a Gaussian distribution for noise. Since many simultaneous tests increase the chances of false positives (the look-elsewhere effect). we adjust to global $p$-value, $p_G$, using Sidak's correction~\cite{Sidak01061967,rug01:000768566}:
\begin{equation}
p_G = 1 - (1 - p_L)^{N_{\rm trials}} \approx N_{\rm trials}\, p_L \quad \text{for } p_L \ll 1.
\end{equation}
where $N_{\rm trials}$ is the number of frequency-pixel pairs. We reject the null hypothesis if $p_G<5$\%. 

\subsection{Follow-up Candidates Identification}

After assessing the significance of our data, we identify sub-threshold candidates for follow-up using more sensitive methods, such as matched-filtering-based searches~\cite{Knee:2023toa}. We first determine the sky pixel with the maximum \ac{SNR}, 
\begin{equation}
\hat{\rho}_{\mathrm{max}}(f) \equiv \mathrm{max}_{\hat{n}}\, \hat{\rho}_{\hat{n}}(f)\,,
\end{equation}
for each frequency bin in both zero-lag and time-shifted data. The full frequency range is divided into 10 Hz sub-bands for the time-shifted data. For each sub-band, we compute the histogram of $\hat{\rho}_{\mathrm{max}}(f)$ and determine the threshold below which $99\%$ of the histogram area lies. This results in an array of sub-band-wise thresholds, which are then smoothed using a running average. Candidates in the zero-lag data with \ac{SNR} exceeding this threshold are selected for further investigation. If the zero-lag data is consistent with random-timeshifted data (and Gaussian noise), this procedure is expected to yield at least $\approx$ 513 candidates. The following calculation shows why this number of candidates is expected:
\begin{equation}
\begin{aligned}
  \!\text{ Sub-bands}
    &= \frac{(1726 - 20)\,\mathrm{Hz}}{10\,\mathrm{Hz}} \\
    &\approx 171,\\
   \text{ Bins per sub-band}&=\frac{10\,\mathrm{Hz}}{\Delta f}=320 \,, \\
  \!\text{ Top-1\% candidates per sub-band}
    &= 320 \times 0.01 \approx 3\,,\\
\!\text{ Total Top-1\% candidates}
    &= 171 \times 3 = 513\,.\\    
\end{aligned}
\end{equation}

\begin{figure*}
    \centering
    \includegraphics[scale=0.45]{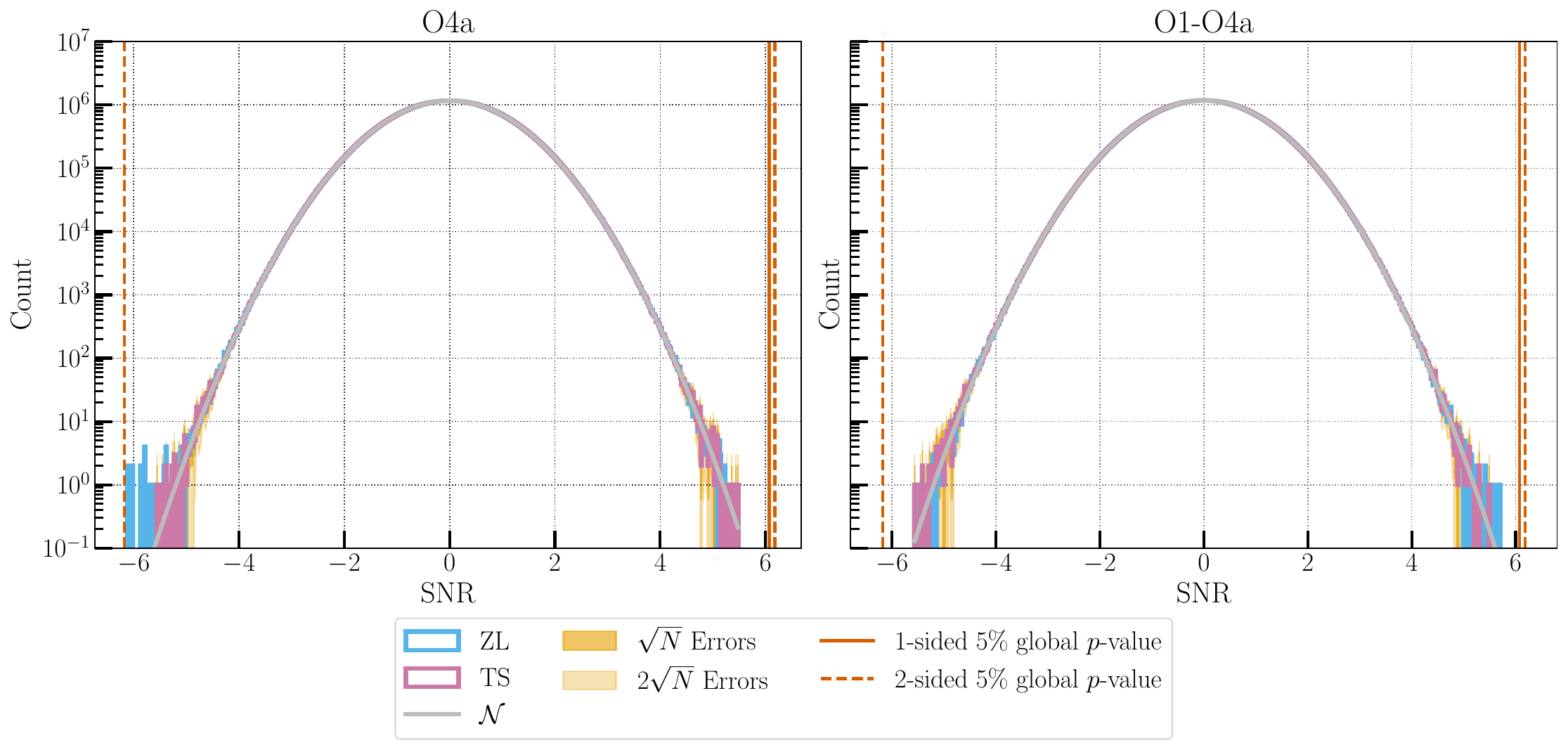}
    \caption{ASAF search results --- Significance: Left Panel: Distribution of \ac{SNR} for the zero-lag (blue) and time-shifted (magenta) datasets with \ac{O4a} observing run. The time-shifted (unphysical) distribution is consistent with a Gaussian (gray line) with mean $-5\times10^{-4}$ and standard deviation 0.98. Poisson 1- and 2-sigma uncertainties for the time-shifted histogram are shown as a yellow-shaded region. The zero-lag histogram is largely consistent with the time-shifted data, except for non-Gaussian excess at negative \ac{SNR}. Vertical brown lines (solid and dashed) indicate the 1- and 2-sided $5\%$ global $p$-value \ac{SNR} thresholds. Although an astrophysical signal is expected to yield a positive \ac{SNR}, we investigate the origin of the observed negative \ac{SNR} feature. Right Panel: Same as the figure in the right panel but with \ac{O1}-\ac{O4a} data.}
    \label{fig:ASAF_SNRhist}
\end{figure*}

\begin{figure*}
    \centering
    \includegraphics[scale=0.45]{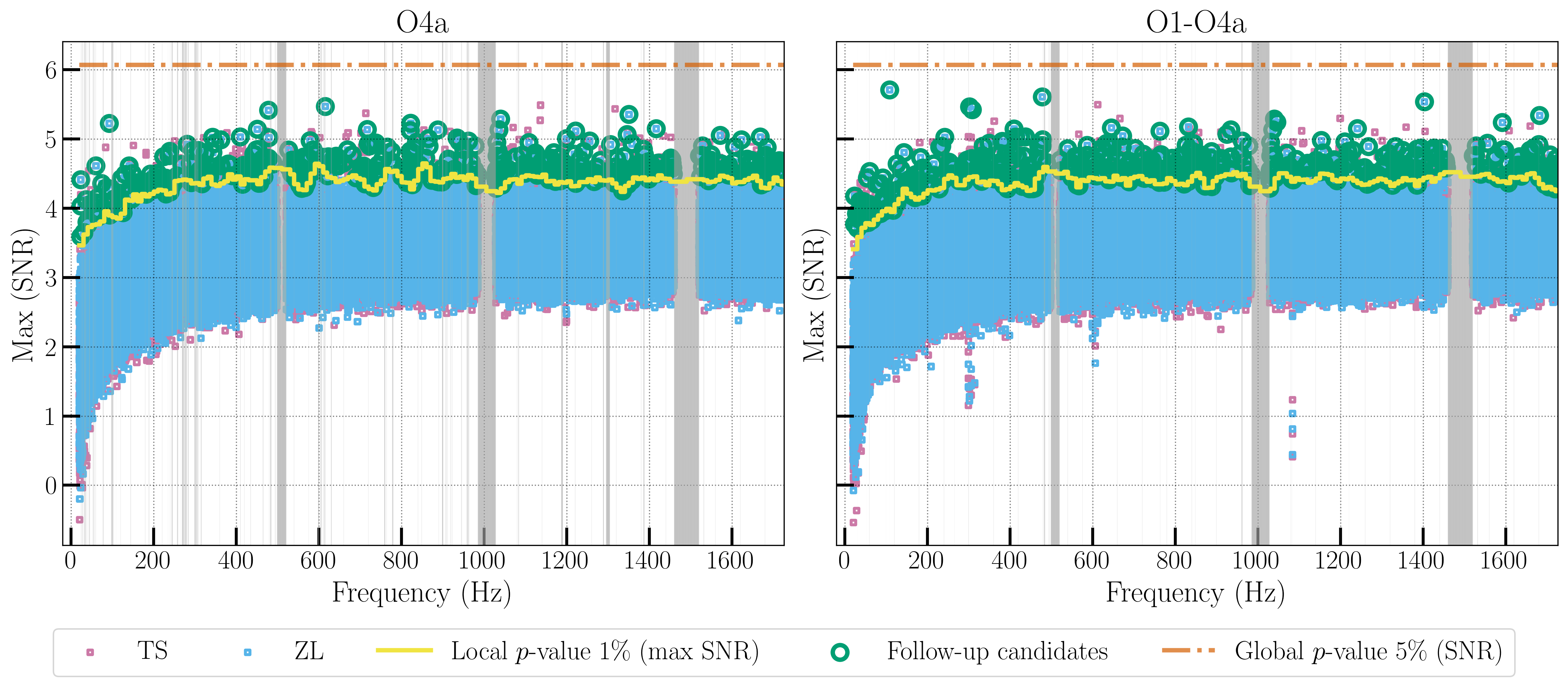}
    \caption{Follow-up sub-threshold candidates: Left Panel: Distribution of the maximum \ac{SNR} statistic (y-axis) as a function of frequency (x-axis) obtained with \ac{O4a} observing run, where Max (SNR) denotes the maximum \ac{SNR} across the sky within each frequency bin. Scatter points for the zero-lag and time-shifted datasets are shown in blue and magenta, respectively. Gray vertical lines indicate vetoed frequency bins excluded due to known instrumental artifacts. The brown horizontal line indicates the \ac{SNR} threshold corresponding to a $5\%$ 1-sided global $p$-value.
    The yellow curve shows the maxSNR threshold corresponding to a $99\%$ local $p$-value in each 10-Hz band, smoothed over three neighboring bands. While the zero-lag data is consistent with Gaussian noise, we identify 505 sub-threshold follow-up candidates marked with teal circles, which may be further analysed using matched-filtering-based \ac{CGW} search pipelines. Right Panel: Same as the left panel, but using data from \ac{O1}-\ac{O4a} observing runs. 
    }
    \label{fig:ASAF_freqMaxSNR}
\end{figure*}

\begin{figure*}
    \centering
    \begin{minipage}{0.48\textwidth}
    \includegraphics[scale=0.425]{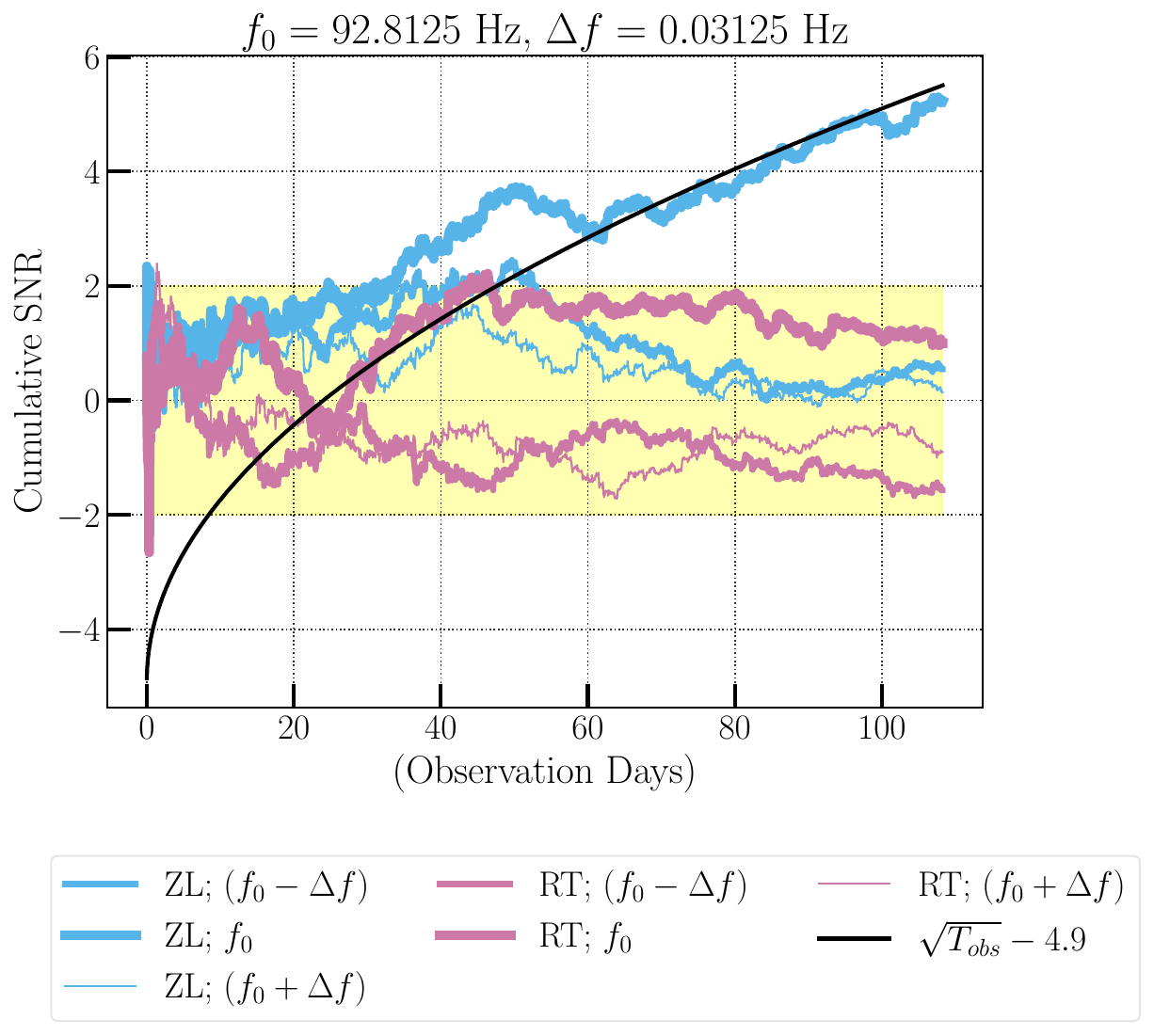}
    \end{minipage}
    \hspace{10pt}
    \begin{minipage}{0.48\textwidth}
    \raggedright
    \includegraphics[scale=0.3]{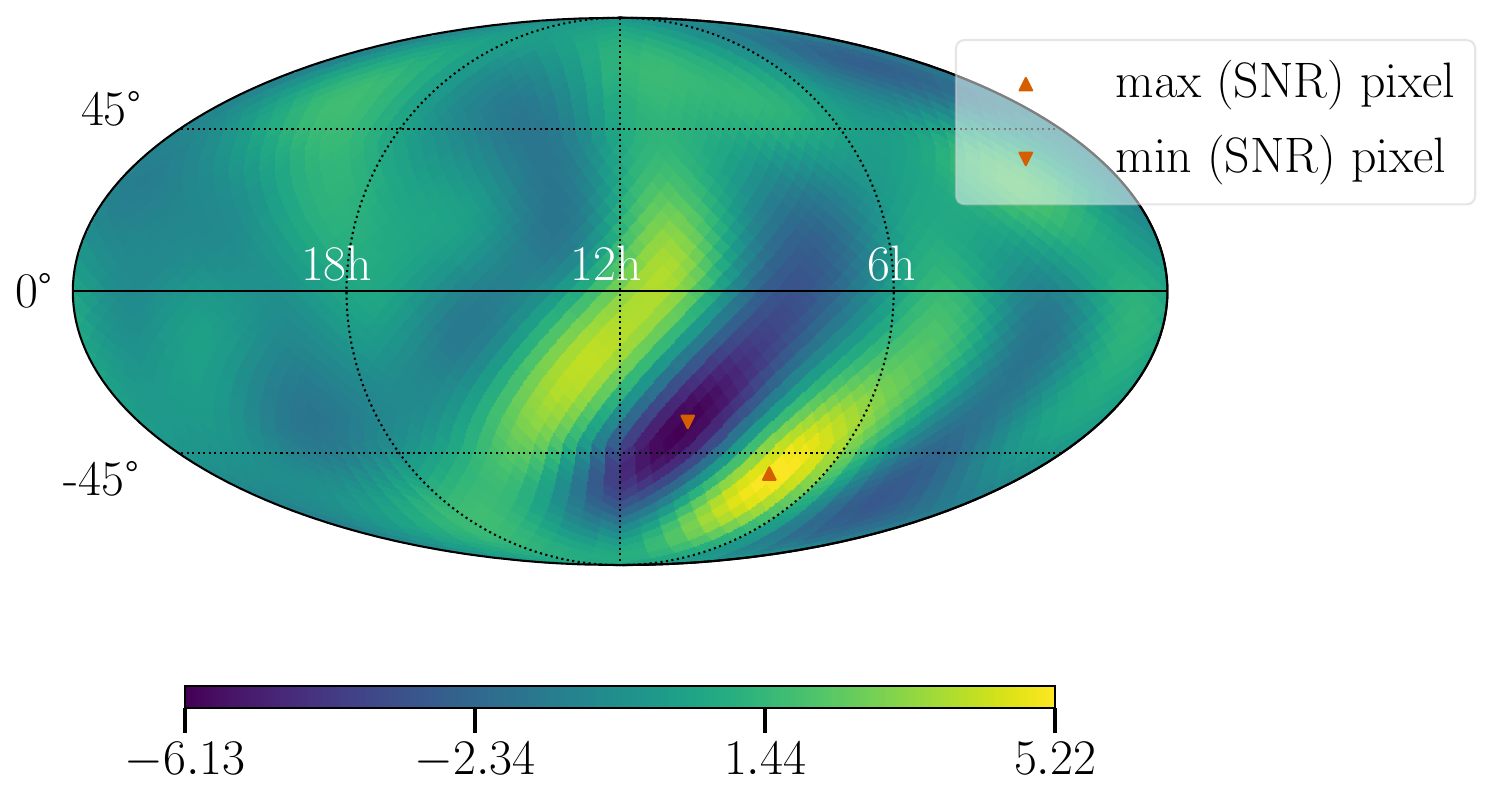}\\
    \includegraphics[scale=0.3]{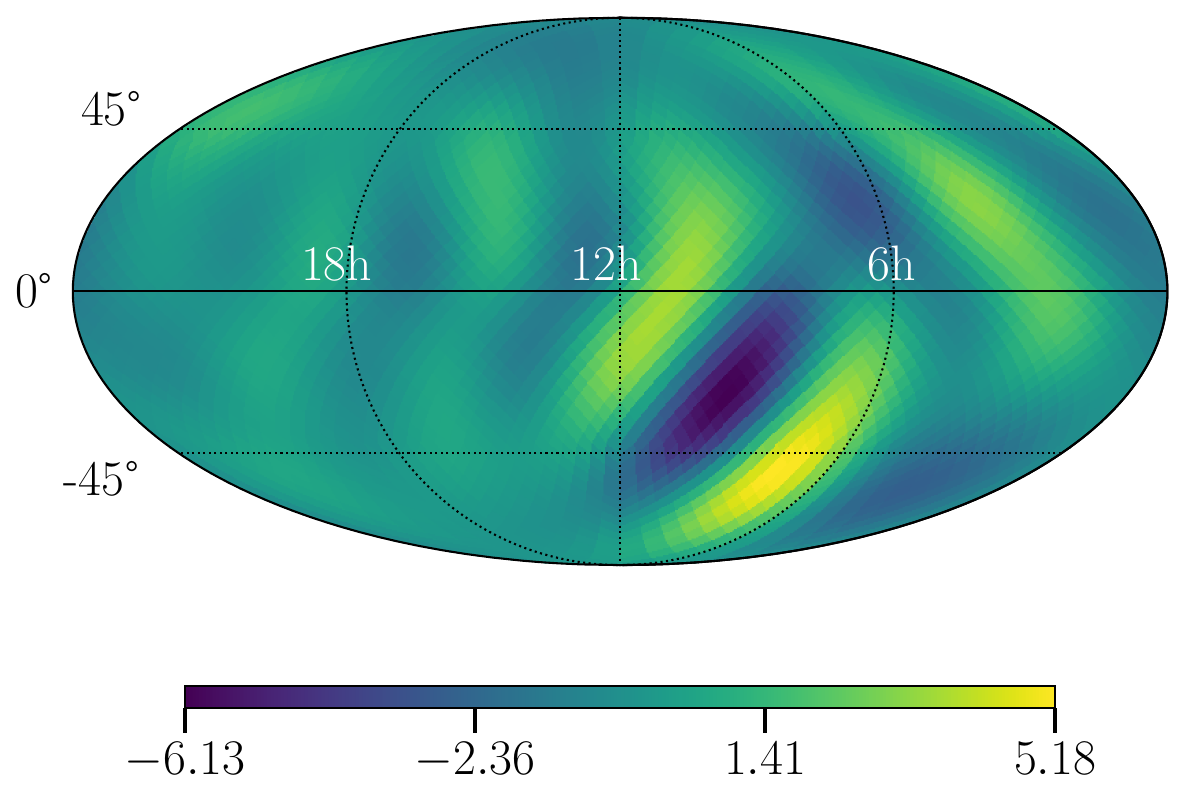}
    \end{minipage}
    \caption{Details of O4a marginal outlier towards negative SNR tail: The SNR skymap for 92.8125 Hz marginal outlier in O4a is shown in the top right panel. The positive and negative \ac{SNR} blobs may be correlated through the point spread function: a \ac{GW} source at the location of maximum \ac{SNR} could produce a corresponding negative \ac{SNR} feature next to it. To test this, we simulate skymap by injecting a source in the positive \ac{SNR} direction and recover a map similar to the observed map, as shown in the bottom right panel. This motivates further investigation of both blobs. In the left panel, we show the evolution of cumulative \ac{SNR} in the direction of maximum SNR (marked by an up triangle in the top right panel) as a function of observing days during O4a. The thick blue indicates the outlier frequency bin, while the two other thin blue curves represent the adjacent frequency bins immediately before and after the outlier. The magenta lines represent SNR evolution for random timeshifted data.  
    }
    \label{fig:OutlierSkymapandEvolution}
\end{figure*}

\subsection{Upper Limit Calculation}\label{app:ASAF_extra_UL}

We adopt a hybrid frequentist-Bayesian approach for setting constraints~\cite{Matas:2020roi}. Assuming the point estimate is a sufficient statistic for measuring \ac{GW} source properties, we apply Bayes' theorem to construct the posterior distribution.

For clarity, the dependence of estimators $\left[\bm{\hat{\mathcal{P}}}_{\hat{n}}(f),\,\bm{\sigma}_{\hat{n}}(f)\right]$ and strain parameter $h_{{\rm eff},\hat{n}}(f)$ on direction and frequency is assumed to be implicit in the following discussion. Each element in the matrices $\left[\bm{\hat{\mathcal{P}}},\,\bm{\sigma}\right]$ here represents an observation from the individual dataset (e.g., baseline or observing run).
\begin{equation}
\label{eq:ASAF:h0_ULs}
    0.95\, =\,\int^{(h_{\mathrm{eff}})^{95\%}}_0\,\mathrm{d}h_{\mathrm{eff}}\, \frac{L(\bm{\hat{\mathcal{P}}}\,\Delta f|\bm{E},h_{\mathrm{eff}})\,p(h_{\mathrm{eff}})}{\int\,\mathrm{d}h_{\mathrm{eff}}\, L(\bm{\hat{\mathcal{P}}}\,\Delta f|\bm{E},h_{\mathrm{eff}})\,p(h_{\mathrm{eff}})}\,.
\end{equation}

The integrand in the above equation represents the posterior distribution for $h_{\mathrm{eff}}$, where $ L(\bm{\hat{\mathcal{P}}}\,\Delta f|\bm{E},h_{\mathrm{eff}}) $ is the likelihood and $ p(h_{\mathrm{eff}})$ is the prior for $h_{\mathrm{eff}}$. The likelihood $L(\bm{\hat{\mathcal{P}}}\,\Delta f|\bm{E},h_{\mathrm{eff}})$ is modelled as a multivariate Gaussian distribution for the random vector $ \bm{\hat{\mathcal{P}}}\,\Delta f\sim\mathcal{N}[(h_{\mathrm{eff}})^2\,\bm{I},\bm{E}]$ where $\bm{I}$ is a column matrix with unit elements. The prior is assumed to follow a uniform distribution, $ h_{\mathrm{eff}} \sim U[0,10\, \sqrt{\sigma\, \Delta f}\,]$~\cite{KAGRA:2021rmt} where $\sigma^2$ is squared sum of variance from individual dataset, i.e., $\sigma^2 = \sum_i\,\Sigma_{ii}$. We note that the integration in the denominator spans the prior-defined range of $h_{\mathrm{eff}}$.

Given that the strain data is obtained by calibrating the detector response to \ac{GW} strain, our estimator is subject to calibration uncertainties (see \cite{LIGOScientific:2016jlg, LIGOScientific:2019vic, KAGRA:2021kbb, LIGOScientific:2016nwa,LIGOScientific:2019gaw,KAGRA:2021mth,KAGRA:2021rmt,Yousuf:2023nmz} for details). Marginalization over calibration uncertainty introduces an additional covariance term to the noise variance matrix of $ \bm{\hat{\mathcal{P}}} $~\cite{Whelan:2012ur}. The modified covariance matrix $ \bm{E} $ is given by  
\begin{equation}
\bm{E} = \bm{\Sigma} + (h_{\mathrm{eff}})^4\,\bm{D} \,,
\end{equation}
where $ \bm{\Sigma} $ is the noise covariance matrix, defined as $ \Sigma_{ij} = \sigma_i^2\,\delta_{ij}\,\Delta f $, and $ \bm{D} $ accounts for calibration errors. The indices $i,j$ run over the number of analyzed datasets. As an example, for the baselines HL and LV, the $ \bm{D} $ matrix is
\begin{equation}
    \label{eq:BBR_methods_ULs_D_matrix_example}
    D = 
    \begin{pmatrix}
    \epsilon_{H}^{2} + \epsilon_{L}^{2} + \epsilon_{H}^{2}\epsilon_{L}^{2} & \epsilon_{L}^{2} \\
    \epsilon_{L}^{2} & \epsilon_{L}^{2}+\epsilon_{V}^{2} + \epsilon_{L}^{2}\epsilon_{V}^{2}
    \end{pmatrix}\,,
\end{equation}
where $\epsilon_{H}$, $\epsilon_{L}$, and $\epsilon_{V}$ are the amplitude calibration uncertainties of the individual detectors \cite{LIGOScientific:2011yag}.\footnote{The uncertainties adopted for different detectors and observing runs in this study are as follows: 
$\epsilon_{\rm H}^{\mathrm{O1}} = 0.048$,   
$\epsilon_{\rm L}^{\mathrm{O1}} = 0.054$, 
$\epsilon_{\rm H}^{\mathrm{O2}} = 0.026$, 
$\epsilon_{\rm L}^{\mathrm{O2}} = 0.0385$, 
$\epsilon_{\rm H}^{\mathrm{O3}} = 0.0696$, 
$\epsilon_{\rm L}^{\mathrm{O3}} = 0.0637$, 
$\epsilon_{\rm V}^{\mathrm{O3}} = 0.05$, 
$\epsilon_{\rm H}^{\mathrm{O4a}} = 0.0693$, and 
$\epsilon_{\rm L}^{\mathrm{O4a}} = 0.041$.} The off-diagonal elements in the matrix are present only when the datasets analyzed from a given observing run share common detectors.

\section{Targeted narrowband radiometer}
\label{sec:NBR_appendix}
\subsection{Source direction and its relevance}
\label{sec:NBR_appendix_source}
In this section, we list the sources chosen for the targeted search, explaining their relevance and selection criteria. 

{Scorpius X-1:} It is a neutron star in a \ac{LMXB} system, considered one of the most promising targets for \ac{CGW} searches \cite{Piccinni:2022vsd,Wette:2023dom}. In several searches \cite{Zhang:2020rph, Vargas:2023gvd, LIGOScientific:2022enz}, a torque balance is assumed between accretion spin-up from the companion star and angular momentum loss by gravitational emission (see also \cite{Papaloizou:1978wnv, Bildsten:1998ey}), and it has a known position and ephemeris. However, its rotational frequency remains unknown \cite{Galaudage:2021cpl}, and a spin wandering effect is expected, caused by stochastic fluctuations of the accretion rate \cite{Mukherjee:2017qme}. The difficulty in characterizing its rotation makes it an optimal target for an unmodeled narrowband search, like the targeted \ac{NBR}, across a large frequency range.

{Galactic Center:} Identified as the location of the supermassive black hole Sagittarius A*, it has been addressed as the host of multiple potential \acp{GW} sources, making it a target for dedicated searches for persistent signals \cite{Dergachev:2019pgs, Piccinni:2019zub, KAGRA:2021rmt}. Being a potential host of many unresolved sources with unknown parameters, it is another natural target for our search.

{SN 1987A:} A young supernova (SN) remnant, SN 1987A has been identified as a promising target for \ac{GW} searches due to its recent origin~\cite{Sun:2016idj, Chung:2011da}. However, until recently, it was not possible to confirm the existence of a neutron star at its center. This has now been established through electromagnetic emission detected by the James Webb Space Telescope~\cite{Fransson:2024csf}. Although the source parameters remain unknown, this confirmation underscores the relevance of SN 1987A as a target for \ac{GW} searches.

{Terzan 5 and NGC 6397:} Globular clusters are known to contain many neutron stars and can be promising targets for \ac{GW} searches. Among the possible selection criteria, we follow \cite{vitral2021does} and \cite{Vitral:2022apu} that point to Terzan 5 and NGC 6397 as a possible host of a large number of unresolved sources, together with the fact that Terzan 5 alone hosts approximately $18\%$ of the known millisecond pulsars \cite{atnf}. However, other criteria exist (for example, see \cite{LIGOScientific:2016ave}).

\subsection{Bin Combination}
\label{sec:NBR_appendix_bin}

The value of the number $N$ of bins to be combined, for the individual sources analyzed, is determined as follows: for Scorpius X-1, assuming torque balance and hence neglecting steady spin variation during the observing time, it is computed using the evolution predicted by its orbital parameters (see \cite{o1supp}). For the Galactic Center, Terzan 5, and NGC 6397 -- which may host a large number of unknown sources -- we compensate only the Earth's Doppler modulation, fixing $N=10$ for all of them.

The high latitude of SN 1987A causes a negligible Doppler spread due to Earth's motion with the default frequency resolution used ($\Delta f = 1/32~\mathrm{Hz}$); hence, in this case, $N=1$.

However, SN 1987A is a young supernova remnant, and it is expected to have a strong spin-down effect due to rotational energy loss. We did not consider the cumulative effect of different observing runs when establishing the bin combination, as this would require specific techniques to account for the several-month gaps between runs, under various assumptions about the source's spin derivatives (see, for example, the study carried out in \cite{Owen:2022mvu}). 

This leads to \ac{UL}s that are less conservative than those obtained by combining additional bins, under the assumption that the entire signal is confined to a single frequency bin rather than distributed across several.

\subsection{Significance}
\label{sec:NBR_appendix_significance}
For a given target $\hat{n}$, to assess the significance of the \ac{SNR}-frequency data results after the bin combination, a $p$-value is estimated with the following process: 1) a large number ($n\sim 256$) noise-only distributions of $\mathcal{P}_{\hat{n}}(f)$ are generated, drawing for each frequency bin a value from Gaussian distributions with mean 0 and the corresponding $\sigma_f$ obtained from the data; 2) for each realization, the generated $\mathcal{P}_{\hat{n}_{\mathrm{sim}}}(f)$ and the measured $\sigma(f)$ are combined according to the data from the given target, yielding a simulated \ac{SNR} distribution with the same bin combination; 3) the maximum \ac{SNR} from each of the simulated distributions is stored; 4) a range of $p$-values running from $[0,1]$ in $n$ steps are generated and associated with the maximum \ac{SNR}s obtained; 5) via a linear interpolation the data \ac{SNR}s are matched to the \ac{SNR}--$p$-value distribution obtained by the simulations. A frequency bin whose \ac{SNR} corresponds to a $p$-value of less than $5\%$ is considered an outlier that needs to be analyzed more thoroughly.

\subsection{Bayesian posterior and \ac{UL} plots}
\label{sec:NBR_appendix_ul}
In the following, the calculation of the upper limits via the integration of Bayesian posteriors will be broken down. The starting point is the definition of expectation value and variance of the cross-correlation statistics for persistent \ac{GW} signals. They can be written respectively \cite{messenger2010understanding}:
\begin{equation}
    \label{eq:expestimator}
\mu_\mathcal{P} = \frac{\sum_{j} (A^{+^2} F^+_{1j} F^+_{2j} + A^{\times^2} F^\times_{1j} F^\times_{2j}) (F^+_{1j} F^+_{2j} + F^\times_{1j} F^\times_{2j})}{\sum_j (F^+_{1j} F^+_{2j} + F^\times_{1j} F^\times_{2j})^2}.
\end{equation}
and 
\begin{equation}
    \label{eq:expsigma}
\sigma^2_\mathcal{P} = \frac{2P_1 P_2}{T_{\rm coh}^2 \sum_j (F^+_{1j} F^+_{2j} + F^\times_{1j} F^\times_{2j})^2},
\end{equation}
where the sum is over the $M$ segments of the semicoherent search with coherence time $T_{\rm coh}$; the \ac{GW} amplitudes for the two polarizations $A^+$, $A^-$ depend on the source inclination angle $\iota$ and the \ac{GW} strain tensor amplitude $h_0$ \footnote{$A^+ = \frac{1}{2} h_0 (1 + \cos^2 \iota)$ and $A^\times = h_0 \cos\iota$, where $\iota$ is the source inclination angle.}; the antenna patterns for the two polarizations and the two detectors at the time segment $j$, $F^{\{+,\times\}}_{\{1,2\}j}$, depend explicitly on the polarization angle $\psi$; the single-sided power spectral density estimations for the two detectors are $P_{1,2}$. In the case of circularly polarised signal $A^+ = A^\times = h_0$ and $\mu_\mathcal{P} = h^2_0$.

The \acp{UL} are computed by integrating the Bayesian posterior function up to the value $h_0^{UL}$ that returns the chosen $95\%$ confidence:
\begin{equation}
\label{eq:ulint}
0.95 = \int_{0}^{h_0^{UL}} p(h_0|\mathcal{P}, \sigma_\mathcal{P}).
\end{equation}

For a given frequency bin, the posterior comes from the marginalization integral
\begin{equation}
    \label{eq:marg}
    \begin{split}
    p(h_0|\mathcal{P}) \propto & 
    \int_{-1}^{1}\mathrm{d}(\cos\iota)p(\iota) 
    \int_{-\pi/4}^{\pi/4}\dd{\psi} p(\psi)\\
    & \int_{-1}^{3} \dd{\lambda}\:p( \mathcal{P} | h_0,\iota, \psi, \lambda) p(\lambda)
    \end{split}
\end{equation}
over the inclination angle $\iota$, the polarization angle $\psi$\cite{messenger2010understanding} and the calibration factor $\lambda$  \cite{Whelan:2012ur}. 

The priors for the first two integrals---respectively $p(\iota)$ and $p(\psi)$---will be uniform distributions within the integral range, and their values will be absorbed in the $\propto$ symbol. The latter integral is over an unknown correction factor representing the uncertainty in the calibration. We assume for it a Gaussian prior distribution $p(\lambda)$, with $\sigma_\lambda$ as standard deviation, a known parameter called \emph{calibration error}.

The likelihood distribution is
\begin{equation}
    \label{eq:likelihood}
    L(\mathcal{P}|h_0,\iota,\psi, \lambda) = \exp \left[- \frac{1}{2}\left(\frac{\lambda \mathcal{P}-\mu_{\mathcal{P}}}{\lambda\sigma_\mathcal{P}}\right)^2\right],
\end{equation}
where the dependence on $h_0$,$\iota$,$\psi$ lies within $\mu_{\mathcal{P}}$.

The marginalized posterior for a generic polarization will be:
\begin{equation}
    \label{eq:posterior}
    \begin{split}
        p(h_0|\mathcal{P}) \propto & \int_{-1}^{1}\mathrm{d}(\cos\iota)
			 \int_{-\pi/4}^{\pi/4}\dd{\psi}\\
        & \int_{-1}^{3}\dd{\lambda}\: \exp \left[- \frac{1}{2}\left(\frac{\lambda \mathcal{P}-\mu_{\mathcal{P}}}{\lambda\sigma_\mathcal{P}}\right)^2 -\frac{1}{2}\left(\frac{\lambda}{\sigma_\lambda}\right)^2\right].
    \end{split}
\end{equation}

In the case of a circularly polarized signal, the likelihood becomes independent of $\iota$ and $\psi$, and the posterior integral will simply reduce to
\begin{equation}
    \label{eq:postcirc}
    L(h_0|\mathcal{P}) \propto \int_{-1}^{3} \dd{\lambda}\: \exp \left[- \frac{1}{2}\left(\frac{\lambda \mathcal{P}-h_0}{\lambda\sigma_\mathcal{P}}\right)^2 -\frac{1}{2}\left(\frac{\lambda}{\sigma_\lambda}\right)^2\right]\,.
\end{equation}

It has been shown that there is a scale factor of $\sim 2.5$ between \acp{UL} computed with a marginalization for a generic polarization with respect to the circular polarization ones, which depends only on the \ac{SNR} of the search results \cite{messenger2010understanding}. Calculating the integrals in \eqref{eq:posterior} for each frequency bin can consume a large amount of computing resources, hence \acp{UL} for the two cases are simulated only for a small number of \ac{SNR} values between -8 and 8. The ratio between the two simulated distributions will be called ``upper limit ratios".

The final \ac{UL} values are computed under the computationally much simpler hypothesis of circular polarization. Assuming a Gaussian prior for the calibration factor, an analytical solution for the marginalized posterior exists \cite{mandicnote,romanonote} in the form of:
\begin{equation}
\label{eq:finalposterior}
    p(h_0|\mathcal{P}) \propto \frac{1}{\sqrt{2\pi E^{2}}} \exp \left[ -\frac{1}{2}\frac{\left(\mathcal{P} - h_0^2\right)^2}{E^2} \right] \, ,
\end{equation}
where $E \equiv \sigma_{\mathcal{P}} + h_0^4 \sigma_{\lambda}$. By interpolation on the real \ac{SNR} data, the \ac{UL} ratios are applied to reproduce the \acp{UL} for a generic polarization.

For the combination of datasets, we have to treat the combined effect of the calibration error of the different runs, changing how the \acp{UL} are produced. It can be shown that the combined likelihood is the product of the single dataset likelihood~\cite{Whelan:2012ur}:
\begin{equation}
\label{eq:multilike}
    L(\boldsymbol{\mathcal{P}}|h_0, \boldsymbol{\lambda}) = \prod_i L(\mathcal{P}_i|h_0, \lambda_i).
\end{equation}
With $\boldsymbol{\mathcal{P}} = \{\mathcal{P}_1,...,\mathcal{P}_n\}$ we indicate the vector of the detection statistics for a fixed frequency bin and with $\boldsymbol{\lambda} = \{\lambda_1,...,\lambda_n\}$ the vector of the unknown calibration factor we want to marginalize, across the $n$ datasets. The marginalized posterior will, in turn, be

\begin{equation}
\label{eq:multimarg}
    p(h_0|\boldsymbol{\mathcal{P}}) \propto  \int \dd{\boldsymbol{\lambda}}  L(\boldsymbol{\mathcal{P}}|h_0, \boldsymbol{\lambda}) p(\boldsymbol{\lambda}).
\end{equation}

It is important to underline that the detector baseline calibration uncertainty affecting our results is the combination of the single detector calibration uncertainties. This, in general, will produce covariance matrices with off-diagonal elements like in the example in \eqref{eq:BBR_methods_ULs_D_matrix_example} that have to be considered in the integration. If, in turn, we consider the same baseline across different runs, we can consider the single-run calibration uncertainties independently of each other and separate the integrals. In this case as well, the marginalized posterior will be the product of the single dataset posteriors:
\begin{equation}
\label{eq:multipost}
   p(h_0|\boldsymbol{\mathcal{P}}) = \prod_i p(h_0|\mathcal{P}_i).
\end{equation}

\subsection{Comparison between \ac{UL} and torque balance for Scorpius X-1}
\label{sec:NBR_appendix_tb}
An order-of-magnitude estimate of the torque-balance level, often used in Sco X-1 \ac{CGW} searches \cite{KAGRA:2022dqk, LIGOScientific:2022enz} to gauge the astrophysical relevance of the explored parameter space, is given by
\begin{equation}
\label{eq:scox1tb}
    h_{0} \approx 3.4 \times 10^{-26} \left( \frac{f_{0}}{600 \, \text{Hz}} \right)^{-1/2}.
\end{equation}

\begin{figure}
     \centering
     \includegraphics[width = \columnwidth]{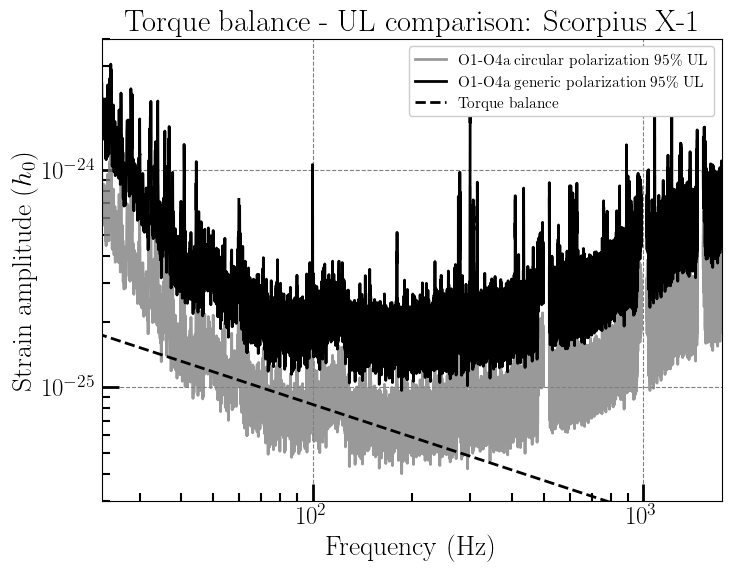}
     \caption{Comparison of the \ac{UL} for generic polarization (solid black line) and a circularly polarized signal (solid gray line) with the torque-balance level from \eqref{eq:scox1tb} (dashed black line). While the former does not reach the torque-balance threshold, the latter lies well below it.}
     \label{fig:scox1tb}
 \end{figure}

This curve is shown in \figref{fig:scox1tb} as a dashed black line. If we compare the $95\%$ confidence level \ac{UL} obtained by the \ac{NBR} search to the torque-balance curve, it is evident that for a generic polarization, the search does not reach amplitudes below the torque balance hypothesis.

On the other hand, if we consider a circularly polarized signal, then the \acp{UL} are not marginalized over inclination and polarization angle, reaching lower effective strain amplitudes. In this scenario, and in agreement with the aforementioned searches, our results surpass the torque-balance limit.
\section{Broadband radiometer search - Statistical significance}
\label{sec:BBR_appendix}
To assess whether a signal is present or not in the BBR search data in \secref{sec:BBR_results}, we consider the \ac{SNR} map for each spectral index and evaluate the statistical significance of the maximum \ac{SNR} under the hypothesis that only noise is present in the data. This is accomplished by simulating $N_{\rm trials}$ realizations of the \ac{SNR} sky-maps and selecting the maximum \ac{SNR} for each of them, hence constructing the probability distribution function of the maximum \ac{SNR}. In practice, for each trial: 1) we simulate a skymap where each pixel contains Gaussian noise; 2) we color the noise in each pixel by using the singular values decomposition of the Fisher matrix to generate the simulated dirty map in the absence of any signal; 3) we use the diagonal of the Fisher matrix 
to obtain $\sigma_{\hat{n}} = ({\rm diag} \left\{\Gamma_{\hat{n} \hat{n}'}\right\})^{-1/2}$ and the simulated estimator map $\hat{\mathcal{P}}_{\hat{n},\,{\rm noise}}$ in the absence of a signal; 4) we obtain the \ac{SNR} map as $\hat{\mathcal{P}}_{\hat{n},\,{\rm noise}} / \sigma_{\hat{n}}$; 5) we select the maximum \ac{SNR} and add it to the histogram of the maximum-\ac{SNR} distribution. After $N_{\rm trials}$, we evaluate the maximum-\ac{SNR} statistical significance ($p$-value) as the ratio of the number of histogram entries that exceed the maximum value of the observed \ac{SNR}s to the total number of the histogram entries (i.e., $N_{\rm trial}$). 

If the statistical significance of the maximum \ac{SNR} is consistent with the noise-only hypothesis, we proceed with the evaluation of the \acp{UL} on the \ac{GWB} angular power spectrum. The evaluation of the \acp{UL} follows the same Bayesian approach as that described in \appref{app:ASAF_extra} by replacing $h_{0}^{\rm eff}$ with $\boldsymbol{\mathcal{P}}$ and $\hat{\boldsymbol{\mathcal{P}}}\,\Delta f$ with $\hat{\boldsymbol{\mathcal{P}}}$ in \eqref{eq:ASAF:h0_ULs}, which we then rescale to \ac{GW} energy flux units.

\section{Spherical harmonics analysis}
\subsection{Angular resolution}
\label{app:SPH_angres}
The number of \ac{SPH} modes to evaluate scales as $(\ell_\mathrm{max} +
1)^2$, and hence, in practice, one cannot search for arbitrary high-order modes
due to substantial computational costs. More importantly, the interferometry in general imposes a
minimum angular scale below which a given baseline becomes insensitive to real signals.
For a monochromatic signal at the frequency $f^*$, the interferometry-limited angular scale is given by
\begin{align}
\label{eq:SPH_method_ang_res}
    \theta=\frac{c}{2\,d\,f^*},
\end{align}
where $d$ is the separation of two \ac{GW} detectors, e.g., $d=3000$~km for the
LIGO detectors. Since \ac{SPH} searches target broadband signals, $f^*$
cannot be uniquely determined, but is approximated by the dominant frequency component of
the signals. This dominant frequency can be estimated by the tangent point
between a given power-law spectrum and the baseline's sensitivity curve, i.e.,
\textit{power-law integrated curve}~\cite{PhysRevD.88.124032}. Therefore, the larger
power-law index $\alpha$ leads to a higher value of $f^*$, e.g., $f^*=52.5, 256.5$~Hz
for $\alpha=0,3$, respectively. Once $\theta$ is known, the highest \ac{SPH}
mode $\lmax$, which corresponds to the minimum angular scale, can be
expressed as $\lmax=\pi/\theta$. Following this argument, we have adopted $\lmax=3,4,16$ for $\alpha=0,2/3,3$,
respectively~\cite{LIGOScientific:2016nwa,LIGOScientific:2019gaw,KAGRA:2021mth}.

\subsection{Regularization of the Fisher matrix}
\label{app:SPH_reg_fisher}
Apart from the angular resolution mentioned above, another numerical issue arises from the fact
that the Fisher matrix is often \textit{ill-conditioned}—that is, some of its
eigenvalues are extremely small, which makes the inversion of the Fisher matrix numerically unstable and introduces additional numerical noise.
This issue can be mitigated by regularizing the Fisher matrix; in previous analyses, we addressed this by discarding negligible eigenmodes.
In \ac{O4a}, we revisited the criterion for identifying such negligible eigenmodes,
taking into account their \ac{RSS} to strike a balance between noise variance and signal recovery bias.
Regularizing the Fisher matrix is
a common mathematical problem, and hence, there are several regularization
methods developed in the field of astronomy~\cite{Eastwood_2018,refId0,10.1111/j.1365-2966.2008.13341.x,PhysRevD.100.043541, PhysRevD.107.122002}. Our approach is to replace the singular values for a subset of
the \ac{SPH} modes (e.g., $M$ out of the total $N$ modes) that have smaller eigenvalues than a given threshold ($\lambda_M$) with infinities, i.e.,

\begin{align}
    \label{eq:SPH_methods_fisher_diag}
\bm{U}\ \bm{\Gamma}\ \bm{V}&=
\mathrm{diag}(\lambda_1, \cdots, \lambda_{M}, \cdots, \lambda_N)\\
&\rightarrow
\mathrm{diag}(\infty , \cdots, \infty, \lambda_{M+1}, \cdots, \lambda_N),
\end{align}
where $\bm{U}, \bm{V}$ are each a unitary matrix, which diagonalizes $\bm{\Gamma}$, and
$\lambda_i$ are the $i$-th singular values of $\bm{\Gamma}$ sorted as
$\lambda_1<\lambda_2<\cdots<\lambda_N$. This would effectively remove the
contribution from the regularized \ac{SPH} modes after inverting the Fisher
matrix~\cite{PhysRevD.80.122002}.
Note that the clean map estimator would be affected by a potential bias in the presence of astrophysical anisotropies 
\begin{align}
    \langle \bm{\phat}_{\mu}\rangle = \bm{\Gamma}^{-1}_\mathrm{R}\cdot
    \bm{\Gamma}\cdot\bm{\mathcal{P}}\neq \bm{\mathcal{P}},
\end{align}
where $\bm{\Gamma}_\mathrm{R}$ is a regularized Fisher matrix. 
Although, technically speaking, the fractional number of singular modes to keep, $f_\mathrm{keep}$, is still arbitrary, optimizing the regularization involves a trade-off relation between a lower variance, i.e.,
larger \ac{SNR}, and the accuracy of the clean map estimator~\cite{PhysRevD.106.023010}.

Since \ac{O1}, \ac{LVK} has adopted an empirical approach that keeps $2/3$ of the total modes, i.e., $f_\mathrm{keep}=(N-M)/N=2/3$ across different $\alpha$ values~\cite{LIGOScientific:2016nwa,LIGOScientific:2019gaw,KAGRA:2021mth}. In \ac{O4a}, we revisit this approach and introduce an alternative justification by computing \ac{RSS} of the clean map estimator for a given simulated signal~\cite{PhysRevD.100.043541, PhysRevD.107.122002}. This is defined as
\begin{align}
    \mathrm{RSS}= |\bm{\phat}_{(k)}-\bm{\mathcal{P}}^\mathrm{inj}|^2,
\end{align}
where $\bm{\mathcal{P}}^\mathrm{inj}$ is an injected \ac{GWB} signal and $\bm{\phat}_{(k)}$ is a clean map estimator with the noise realization using the Fisher matrix derived from the time-shifted \ac{O4a} dataset, and the regularization keeping $k$ singular modes. For each $k$, we repeat this computation with 10 different noise realizations and take their mean value. Since the mean of \ac{RSS} follows
\begin{align}
\langle\mathrm{RSS}\rangle & =\langle| \bm{\phat}_{(k)}|^2\rangle-2 \Re\left[\langle\bm{\phat}_{(k)}\rangle\cdot\bm{\calP}^\mathrm{inj}\right]+|\bm{\calP}^\mathrm{inj}|^2 \\
\label{eq:SPH_methods_rss_var_bias}
& =\underbrace{\sigma_\mathcal{P}^2}_\mathrm{variance} + \underbrace{\left|\langle\bm{\phat}_{(k)}\rangle-\bm{\calP}^\mathrm{inj}\right|^2}_\mathrm{bias},
\end{align}
the first and second terms in \eqref{eq:SPH_methods_rss_var_bias} represent the variance and bias of the clean map, respectively. Therefore, minimizing $\langle\mathrm{RSS}\rangle$ in terms of $k$ identifies the optimal point that compromises the variance and bias. In principle, \ac{RSS} values depend on the anisotropy model of the injected \ac{GWB}, and here we consider only the monopole component, i.e., $\mathcal{P}_{\ell m}=0$ ($\ell\neq0$ or $m\neq0$), as a plausible detection scenario.

\begin{figure}
\centering
\includegraphics[width=\columnwidth]{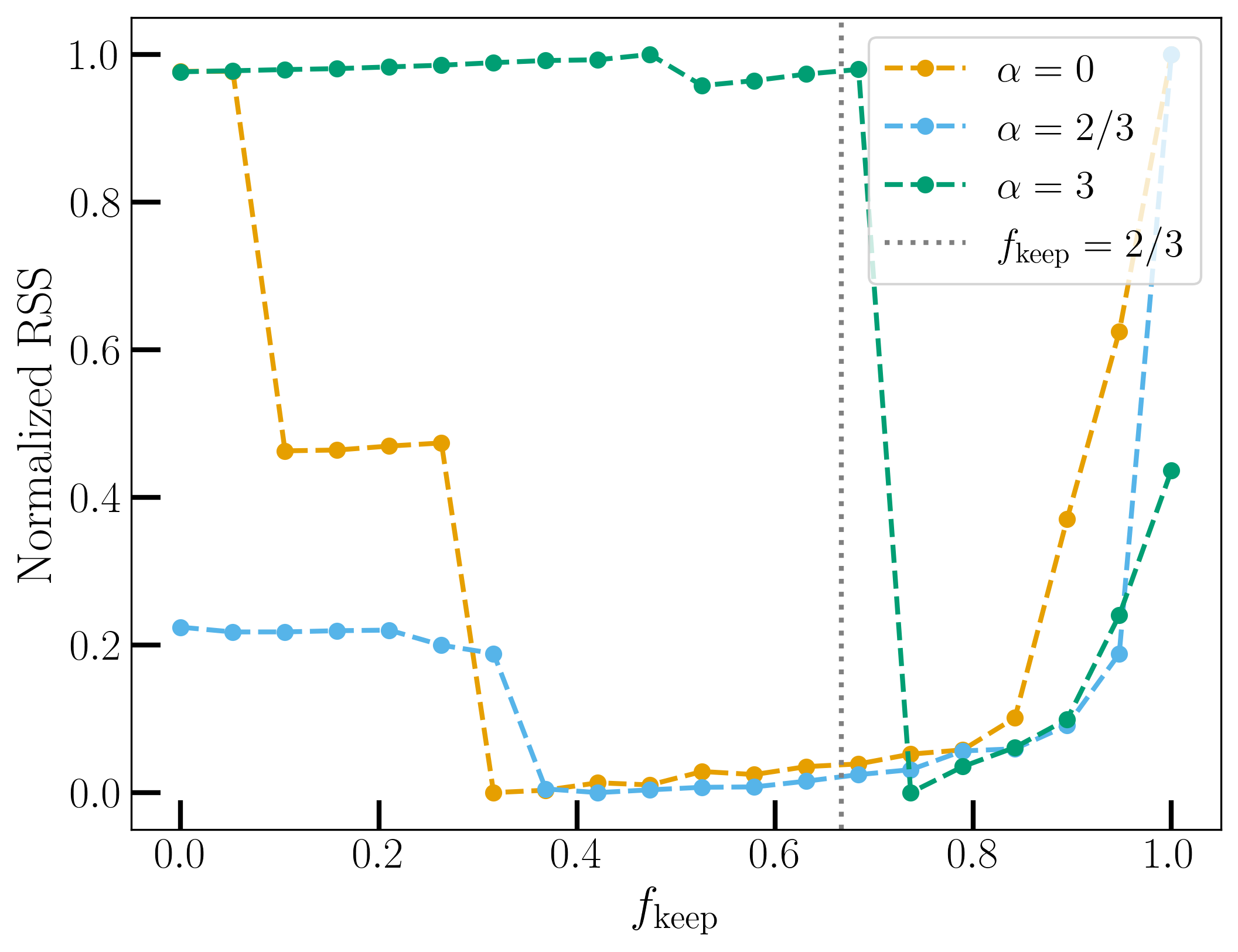}
\caption{Normalized \ac{RSS} as a function of $f_\mathrm{keep}$ for each $\alpha$ value.}
\label{fig:RSS_alphas}
\end{figure}

\figref{fig:RSS_alphas} shows a \textit{normalized} \ac{RSS} as a function of $f_\mathrm{keep}$ for each $\alpha$ value, which demonstrates the optimal $f_\mathrm{keep}$ of 0.3, 0.35 and 0.72 for $\alpha=0,2/3,3$, respectively. We will adopt these values for the regularization applied to the main results of the \ac{SPH} analysis.

\begin{figure}
\centering
\includegraphics[width=\columnwidth]{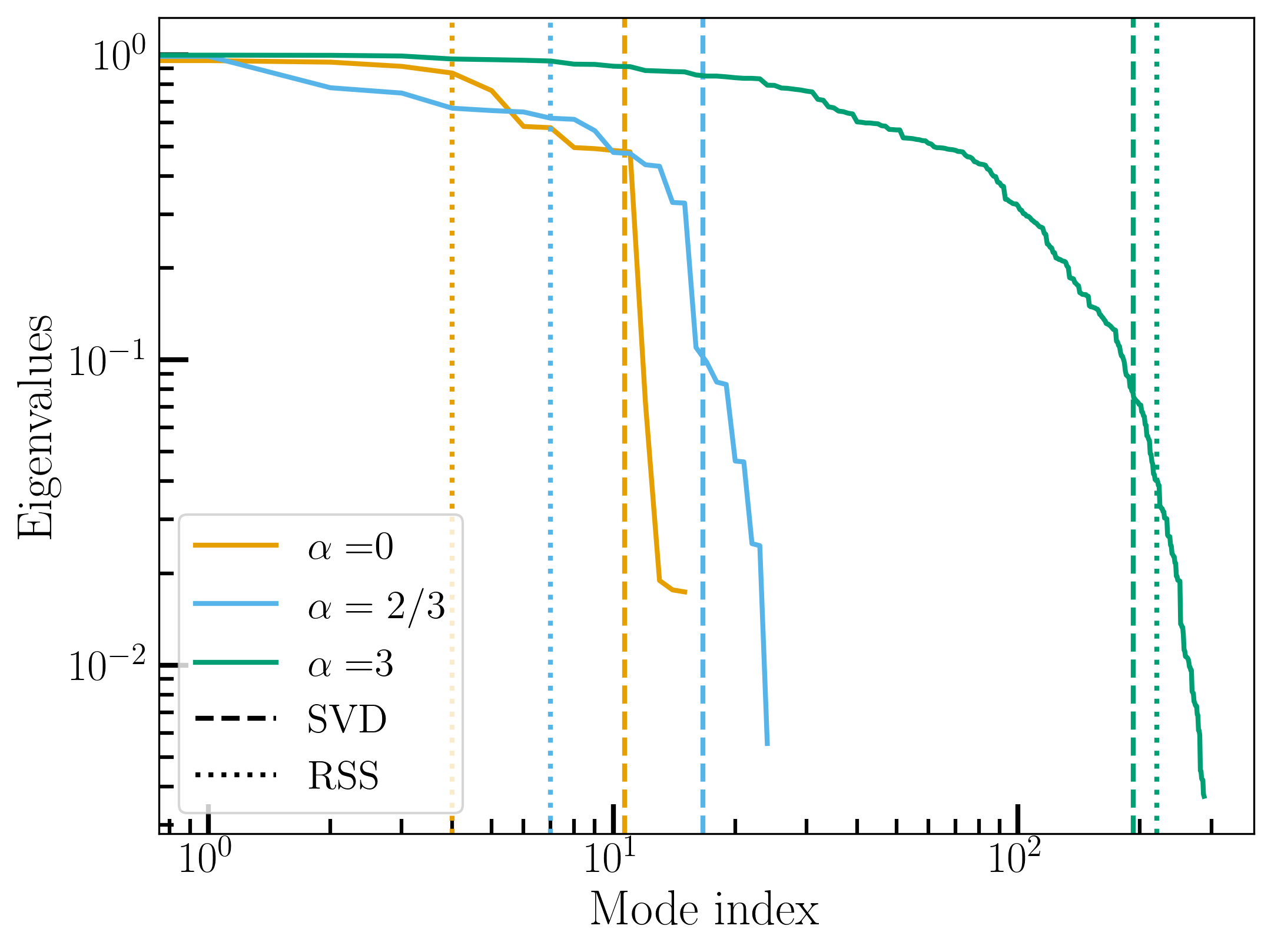}
\caption{Distribution of eigenvalues for the
Fisher matrix with different $\alpha$ values combining \ac{O1} to \ac{O4a} data. The two dashed lines with each color compare the number of modes to keep between the conventional ($f_\mathrm{keep}=2/3$) and new (RSS) regularization methods.}
\label{fig:fisher_condition}
\end{figure}

Also, \figref{fig:fisher_condition} shows the distribution of eigenvalues for the
Fisher matrix with different $\alpha$ values combining \ac{O1} to \ac{O4a} data. The two dashed lines with each color compare the number of modes to keep between the conventional ($f_\mathrm{keep}=2/3$) and the \ac{RSS} regularization methods. Compared to the conventional $f_\mathrm{keep}=2/3$, the new regularization keeps fewer modes for $\alpha=0, 2/3$, while keeping more modes for $\alpha=3$. Although this change in $f_\mathrm{keep}$ alters the variance of the clean map estimator, we note that it does not indicate a physical change in the sensitivity, but rather a different way of presenting the results.

\subsection{Cross $\hat{C}_\ell$ estimator}
\label{app:SPH_cross_Cell}
In the limiting case where the detector data contain only detector noise, the
auto-${\hat{C}_\ell}$ estimator given by \eqref{eq:SPH_methods_curr_estimator}
is unbiased. However, Ref.~\cite{PhysRevD.109.103535} showed that in the
presence of an astrophysical \ac{GWB}, it is significantly biased by the
shot-noise-dominated nature of the signal, which arises from the discrete spatial or temporal realization of individual events. Also, this shot noise effect scales as
$\propto1/T_\mathrm{obs}$ in terms of $C_\ell$, where ${T_\mathrm{obs}}$ is the
observation time, which is the same scaling expected for the \acp{UL} set by
the search. As a result, shot noise may become a limiting factor in future \ac{SPH} searches—particularly if sensitivity improves faster than the $1/T_\mathrm{obs}$ scaling, whether due to enhanced detector sensitivity or the inclusion of additional detectors.  In such a regime, the analysis could begin to detect \ac{GWB} signals from plausible astrophysical populations, e.g., from \acp{CBC}. In addition, Ref.~\cite{PhysRevD.109.103535} found that the auto-${\hat{C}_\ell}$ estimator is biased even in the absence of shot noise in the GWB signal.

Therefore, we introduced the cross-${\hat{C}_\ell}$ estimator shown in \eqref{eq:SPH_methods_opt_estimator}, which involves multiple clean maps derived from subsets of the whole dataset. For each subset, we compute the corresponding dirty map $\bm{X}^{(i)}$ and Fisher matrix $\bm{\Gamma}^{(i)}$, 
and follow the same deconvolution procedure as described earlier to obtain the individual clean maps.
This cross-$\hat{C}_\ell$ estimator, obtained by
summing over all pairs of distinct maps, is by construction unbiased, accounting for the fact that the bias arises from autocorrelation of each data segment. This concerns contributions to the bias not only due to the GWB signal and shot noise but also the detector noise, which is why $\bm{\Sigma}_{\mathrm{R}}$
is no longer subtracted from the clean map product, unlike in \eqref{eq:SPH_methods_curr_estimator}.
In practice, there are many ways to divide the total dataset into subsets, with no trivial choice. However, following Ref.~\cite{PhysRevD.109.103535}, we aim for constructing subsets of the data with approximately equal sensitivity, leading to the best \ac{UL} on the $C_\ell$ measurement. Therefore, in our case, we combine data across different observing runs and baselines to yield data subsets with approximately equal sensitivity as follows:
The data from the \ac{O1} and \ac{O2}, as well as the HV and LV baselines of the \ac{O3}, are combined into one subset. The total dirty map and Fisher matrix of this subset are then given by
\begin{align}
\begin{split}
    \bm{X}^{(1)} &= \bm{X}^\mathrm{O1} + \bm{X}^\mathrm{O2} + \bm{X}^\mathrm{O3(HV)} + \bm{X}^\mathrm{O3(LV)}\, ,\\
    \bm{\Gamma}^{(1)} &= \bm{\Gamma}^\mathrm{O1} + \bm{\Gamma}^\mathrm{O2} + \bm{\Gamma}^\mathrm{O3(HV)} + \bm{\Gamma}^\mathrm{O3(LV)}\, ,
\end{split}
\end{align}
respectively, similar to Ref.~\cite{LIGOScientific:2019gaw}. We construct the first subset in this manner and divide the HL baseline dataset from \ac{O3} and \ac{O4a} according to the sensitivity of this dataset, since \ac{O1}, \ac{O2}, and the HV and LV baselines of \ac{O3} are significantly less sensitive than the HL baselines of \ac{O3} and \ac{O4a}. Specifically, the data from the HL baseline of \ac{O3} and \ac{O4a} are divided equally into 6 and 11 subsets, respectively. For each subset, we compute the corresponding dirty map, $\bm{X}^{(i)}$, and Fisher matrix, $\bm{\Gamma}^{(i)}$, where the index $i$ ranges from 2 to 7 for \ac{O3} and from 8 to 18 for \ac{O4a}.

\subsection{\ac{UL} and significance computation}
\label{app:SPH_ul_pvalue}
\begin{figure*}
\centering
\includegraphics[width=\textwidth]{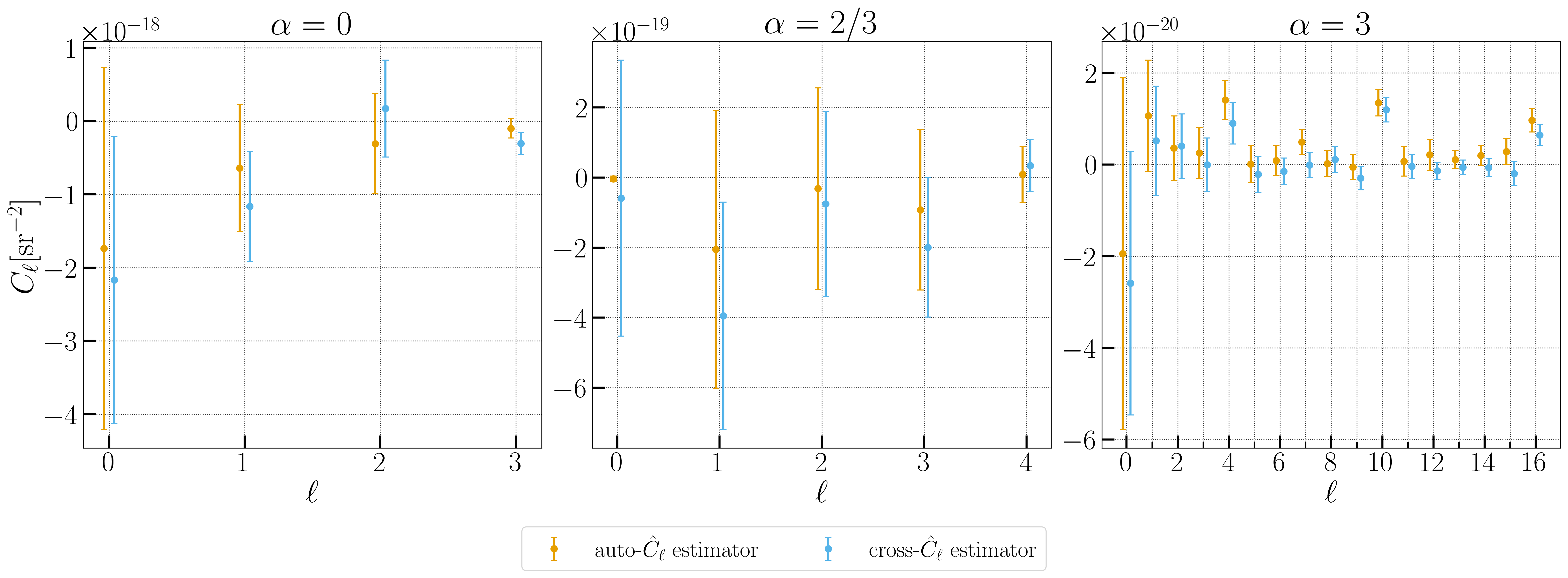}
\caption{$C_\ell$ point estimates using either type of the estimators mentioned above for each power-law spectrum model with the combined O1+O2+O3+O4a data.}
\label{fig:Cl_pt_est}
\end{figure*}

Following the procedure for significance assessment described in \secref{sec:sph_methods}, we compute the $p$-value for each real and imaginary part of the clean map estimator for the combined dataset across \ac{O1}-\ac{O4a}, and compare all these $p$-values to the local and global $p$-value thresholds, respectively. For better visualization, we convert each $p$-value to a $z$-score, which ranges between $[-\infty, \infty]$, through the inverse error function.
\begin{figure*}
\centering
\includegraphics[width=\textwidth]{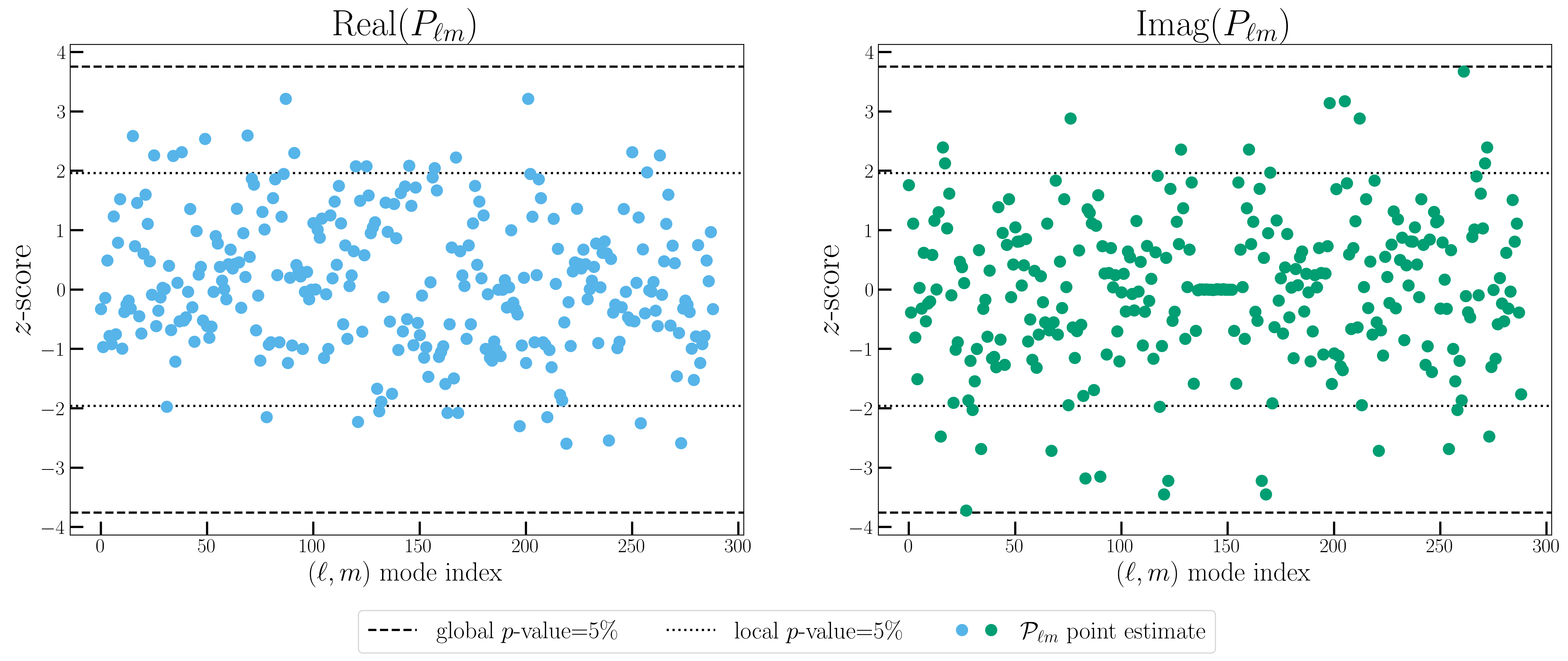}
\caption{$z$-score distribution of $\alpha=3$ case for the combined O1-O4a data.}
\label{fig:zscore_alpha3}
\end{figure*}
Including the $\alpha=3$ case shown in \figref{fig:zscore_alpha3}, all \ac{SPH} modes lie within the global $p$-value threshold (the dashed lines) across the three power-law spectrum models, indicating the consistency with a Gaussian distribution.
Similarly, we investigate $p$-values for the 18 subsets of data used to compute the cross-${\hat{C}_\ell}$ estimator and confirm that they remain within the global $p$-value threshold with very few exceptions\footnote{Considering all individual data subsets and for $\alpha\in\{2/3,3\}$, the global $p$-value threshold was exceeded for a negligible number of SPH modes. For $\alpha = 0$, however, this occurred for approximately 4\% of all SPH modes. Due to this notable deviation from Gaussianity in the latter case, the results of the cross-${\hat{C}_\ell}$ estimator for $\alpha=0$ should be interpreted with caution.}.
Therefore, we confirm that no confident detection is made at a $p$-value below $<5\%$. We then compute the $C_\ell$ point estimates and their 1$\sigma$ uncertainty using \eqsref{eq:SPH_methods_curr_estimator_sigma}{eq:SPH_methods_opt_estimator_sigma} for either of the two estimators, respectively. 
 Although the auto-$\hat{C}_\ell$ estimator is potentially biased due to shot noise in an astrophysical \ac{GWB} signal, \figref{fig:Cl_pt_est} shows that the differences between the two estimators remain consistently within their 1$\sigma$ error bars across all $\ell$ values. This suggests that such a bias is not noticeable, as expected in the absence of a detected signal, given the current search sensitivity.

Given the absence of a detected \ac{GWB} signal, we compute \acp{UL} on
$C_\ell^{1/2}$ at each angular scale characterized by the $\ell$ value, for
different power-law frequency spectrum models, by constructing Bayesian
posteriors from the Monte Carlo samples using the same procedure described for
computing $p$-values in \secref{sec:sph_methods}. The $\bm{\phat}^\mathrm{sim}$
samples generated for the whole dataset and each of the 18 subsets are converted
into the auto-${\hat{C}_\ell}$ and cross-${\hat{C}_\ell}$ estimators based on
\eqref{eq:SPH_methods_curr_estimator} and \eqref{eq:SPH_methods_opt_estimator},
respectively. We add the observed $C_\ell$ point estimates to each
sample and construct the simulated \ac{PDF}, $p(C_\ell | \hat{C}_\ell)$.
Furthermore, we account for the calibration uncertainty by marginalizing this
simulated \ac{PDF} over the calibration factor $\lambda$ such that
\begin{align}
    p(C_\ell | \hat{C}_\ell) = \int \mathrm{d}\lambda\, p(C_\ell | \hat{C}_\ell, \lambda)\, p(\lambda) \, ,
\end{align}
where
\begin{align}
    p(C_\ell | \hat{C}_\ell, \lambda) = p(C_\ell/\lambda^2| \hat{C}_\ell, \lambda_0=1)\, ,\\
    p(\lambda) \propto\exp{\left(-\frac{(\lambda-1)^2}{2\sigma_\lambda^2}\right)}\, .
\end{align}
The variance of the calibration factor, $\sigma^2_\lambda$, is given by the sum of the individual calibration uncertainties measured in each baseline and observing run\footnote{Technically, for the cross-${\hat{C}_\ell}$ estimator, the propagation of the calibration error differs from the auto-${\hat{C}_\ell}$ estimator because it computes the cross-power of the clean map across the subsets rather than its auto-power, and hence this calibration factor should be derived differently. We leave this modification as future work.}, i.e., $\sigma^2_\lambda=\sum_i\epsilon_i^2$.
Eventually, we identify 95th percentiles of $p(C_\ell | \hat{C}_\ell)$ and define it as the \ac{UL} at the 95\% confidence level.

When we generate the $\bm{\phat}^\mathrm{sim}$
samples, we need to account for a complication due to the requirement that the clean maps be real quantities on the pixel basis; each sample needs to conserve the following symmetry
\begin{align}
    \Gamma_{\ell -m, \ell^\prime-m^\prime}=(-1)^{m+m^\prime}\Gamma_{\ell m, \ell^\prime m^\prime}^*\, .
\end{align}
The clean map estimator $\phat_{\ell m}$ follows a multi-variate \textit{complex} Gaussian distribution with zero mean and the covariance matrix $\bm{\Sigma}$ given by \eqref{eq:SPH_methods_sigmas_curr_C_l_estimator}. To satisfy this condition, we draw each real and imaginary part of a sample separately by constructing a \textit{real} covariance matrix
\begin{align}
    \binom{\mathrm{Re}\bm{\phat}^\mathrm{sim}}{\mathrm{Im}\bm{\phat}^\mathrm{sim}} \sim \mathcal{N}\left(\left[\begin{array}{l}
0 \\
0
\end{array}\right], \left[\begin{array}{cc}
\bm{\Sigma}_\mathrm{RR} & \bm{\Sigma}_\mathrm{RI} \\
\bm{\Sigma}_\mathrm{IR} & \bm{\Sigma}_\mathrm{II}
\end{array}\right]\right)\, ,
\end{align}
where each block matrix in the real covariance matrix is defined as
\begin{align}
&
    \begin{aligned}
        \bm{\Sigma}_\mathrm{RR} &= \frac{1}{4}\{\Sigma_{l m l^{\prime} m^{\prime}}+(-1)^m \Sigma_{l-m l^{\prime} m^{\prime}}\\
        &+(-1)^{m^{\prime}} \Sigma_{l m l^{\prime}-m^{\prime}}+(-1)^{m+m^{\prime}} \Sigma_{l-m l^{\prime}-m^{\prime}}\} \,,
    \end{aligned}\\
    &
    \begin{aligned}
    \bm{\Sigma}_\mathrm{II} &=\frac{1}{4}\{\Sigma_{l m l^{\prime} m^{\prime}}-(-1)^m \Sigma_{l-m l^{\prime} m^{\prime}}\\
    &-(-1)^{m^{\prime}} \Sigma_{l m l^{\prime}-m^{\prime}}+(-1)^{m+m^{\prime}} \Sigma_{l-m l^{\prime}-m^{\prime}}\} \,,
    \end{aligned}\\
    &
    \begin{aligned}
        \bm{\Sigma}_\mathrm{RI} &=\bm{\Sigma}_\mathrm{IR}^\top\\
        &= \frac{-i}{4}\{\Sigma_{l m l^{\prime} m^{\prime}}+(-1)^m \Sigma_{l-m l^{\prime} m^{\prime}}\\
        &-(-1)^{m^{\prime}} \Sigma_{l m l^{\prime}-m^{\prime}}-(-1)^{m+m^{\prime}} \Sigma_{l-m l^{\prime}-m^{\prime}}\} \, .\\
    \end{aligned}
\end{align}
\bibliography{ref_list}

\end{document}